\documentclass[sigconf]{acmart}

\AtBeginDocument{%
  }

\copyrightyear{2025}
\acmYear{2025}
\setcopyright{acmlicensed}
\acmConference[WSDM '25] {Proceedings of the Eighteenth ACM International Conference on Web Search and Data Mining}{March 10--14, 2025}{Hannover, Germany.}
\acmBooktitle{Proceedings of the Eighteenth ACM International Conference on Web Search and Data Mining (WSDM '25), March 10--14, 2025, Hannover, Germany}
\acmISBN{979-8-4007-1329-3/25/03}
\acmDOI{10.1145/3701551.3703579}

\usepackage[utf8]{inputenc} % allow utf-8 input
\usepackage[T1]{fontenc}    % use 8-bit T1 fonts
\usepackage{hyperref}       % hyperlinks
\usepackage{url}            % simple URL typesetting
\usepackage{booktabs}       % professional-quality tables
\usepackage{amsfonts}       % blackboard math symbols
\usepackage{nicefrac}       % compact symbols for 1/2, etc.
\usepackage{microtype}      % microtypography
\usepackage{xcolor}         % colors
\usepackage{wrapfig}

\usepackage[ruled,vlined]{algorithm2e}
\usepackage{graphicx} 
\usepackage{amsmath}
\usepackage{amsthm}
\usepackage{subfigure}
\usepackage{array} 
\usepackage{multirow}
\usepackage{color}
\usepackage{adjustbox}

\newcommand{\ie}{\emph{i.e., }}
\newcommand{\eg}{\emph{e.g., }}

\newcommand{\cf}{\emph{cf. }}
\newcommand{\aka}{\emph{a.k.a. }}

\newtheorem{theorem}{Theorem}
\newtheorem{observation}{Observation}

\begin{document}

%%
%% The "title" command has an optional parameter,
%% allowing the author to define a "short title" to be used in page headers.

\title{How Do Recommendation Models Amplify Popularity Bias?  An Analysis from the Spectral Perspective}

\author{Siyi Lin}
\affiliation{%
  \institution{Zhejiang University}
  \city{Hangzhou}
  \country{China}
}
\email{lin452_lsy@zju.edu.cn}
\authornotemark[2]
\authornotemark[3]

\author{Chongming Gao}
\affiliation{%
  \institution{University of Science and Technology of China}
  \city{Hefei}
  \country{China}
}
\email{chongming.gao@gmail.com}

\author{Jiawei Chen}
\affiliation{%
  \institution{Zhejiang University}
  \city{Hangzhou}
  \country{China}
}
\email{sleepyhunt@zju.edu.cn}
\authornote{Corresponding author.}
\authornote{State Key Laboratory of Blockchain and Data Security, Zhejiang University.}
\authornote{College of Computer Science and Technology, Zhejiang University.}
\authornote{Hangzhou High-Tech Zone (Binjiang) Institute of Blockchain and Data Security.}

\author{Sheng Zhou}
\affiliation{%
  \institution{Zhejiang University}
  \city{Hangzhou}
  \country{China}
}
\email{zhousheng_zju@zju.edu.cn}

\author{Binbin Hu}
\affiliation{%
  \institution{Ant Group}
  \city{Hangzhou}
  \country{China}
}
\email{bin.hbb@antfin.com}

\author{Yan Feng}
\affiliation{%
  \institution{Zhejiang University}
  \city{Hangzhou}
  \country{China}
}
\email{fengyan@zju.edu.cn}
\authornotemark[2]
\authornotemark[3]

\author{Chun Chen}
\affiliation{%
  \institution{Zhejiang University}
  \city{Hangzhou}
  \country{China}
}\email{chenc@zju.edu.cn}
\authornotemark[2]
\authornotemark[3]

\author{Can Wang}
\affiliation{%
  \institution{Zhejiang University}
  \city{Hangzhou}
  \country{China}
}\email{wcan@zju.edu.cn}
\authornotemark[2]
\authornotemark[4]
\renewcommand{\shortauthors}{Siyi Lin et al.}

\begin{abstract}
Recommendation Systems (RS) are often plagued by popularity bias. When training a recommendation model on a typically long-tailed dataset, the model tends to not only inherit this bias but often exacerbate it, resulting in over-representation of popular items in the recommendation lists. This study conducts comprehensive empirical and theoretical analyses to expose the root causes of this phenomenon, yielding two core insights: 1) Item popularity is memorized in the principal spectrum of the score matrix predicted by the recommendation model; 2) The \textit{dimension reduction} phenomenon amplifies the relative prominence of the principal spectrum, thereby intensifying the popularity bias.

Building on these insights, we propose a novel debiasing strategy that leverages a \textit{spectral norm regularizer} to penalize the magnitude of the principal singular value. We have developed an efficient algorithm to expedite the calculation of the spectral norm by exploiting the spectral property of the score matrix. Extensive experiments across seven real-world datasets and three testing paradigms have been conducted to validate the superiority of the proposed method. The code is available at https://github.com/LIN452/ReSN/.

\end{abstract}

%%
%% The code below is generated by the tool at http://dl.acm.org/ccs.cfm.
%% Please copy and paste the code instead of the example below.
%%
\begin{CCSXML}
<ccs2012>
 <concept>
  <concept_id>10010520.10010553.10010562</concept_id>
  <concept_desc>Computer systems organization~Embedded systems</concept_desc>
  <concept_significance>500</concept_significance>
 </concept>
 <concept>
  <concept_id>10010520.10010575.10010755</concept_id>
  <concept_desc>Computer systems organization~Redundancy</concept_desc>
  <concept_significance>300</concept_significance>
 </concept>
 <concept>
  <concept_id>10010520.10010553.10010554</concept_id>
  <concept_desc>Computer systems organization~Robotics</concept_desc>
  <concept_significance>100</concept_significance>
 </concept>
 <concept>
  <concept_id>10003033.10003083.10003095</concept_id>
  <concept_desc>Networks~Network reliability</concept_desc>
  <concept_significance>100</concept_significance>
 </concept>
</ccs2012>
\end{CCSXML}
\ccsdesc[500]{Information systems~Recommender systems}

%%
%% Keywords. The author(s) should pick words that accurately describe
%% the work being presented. Separate the keywords with commas.
\keywords{Recommender System; Popularity Bias}
%% A "teaser" image appears between the author and affiliation
%% information and the body of the document, and typically spans the
%% page.

% \received{20 February 2007}
% \received[revised]{12 March 2009}
% \received[accepted]{5 June 2009}

%%
%% This command processes the author and affiliation and title
%% information and builds the first part of the formatted document.
\maketitle

\section{Introduction}
Recommender Systems (RS), with their capability to offer personalized suggestions, have found applications across various domains \cite{covington2016deep,qiu2018deepinf,zhou2018deep}. Collaborative filtering (CF),  a widely-used technique within RS, learns user preference from historical interactions. However, their effectiveness in personalization is significantly compromised by popularity bias \cite{chen2023bias}. This bias emerges when user interaction data showcases a long-tailed distribution of item interaction frequencies. Subsequently, recommendation models trained on such data tend to inherit and even amplify this bias, leading to an overwhelming presence of popular items in recommendation results \cite{wei2021model,zhang2021causal,zhu2021popularity}. 

% Recommender Systems (RS), with their capability to offer personalized suggestions, have found applications across various domains \cite{covington2016deep,qiu2018deepinf,zhou2018deep}. Nevertheless, their effectiveness in personalization is significantly compromised by popularity bias \cite{chen2023bias}. This bias emerges when recommendation data showcases a long-tailed distribution of item interaction frequencies. Subsequently, recommendation models trained on such data tend to inherit and even amplify this bias, leading to an overwhelming presence of popular items in recommendation results \cite{wei2021model,zhang2021causal,zhu2021popularity}. 
% Figure \ref{fg:pro} illustrates this issue by examining well-known models (\eg MF \cite{hu2008collaborative,he2016fast} and LightGCN \cite{he2020lightgcn}) on benchmark datasets (\eg MovieLens, Douban and Globo). We discover that a mere 0.6\% of the most popular items, accounting for 20\% of total interactions in data, occupy over 63\% recommendation slots in Douban. 
This notorious effect not only undermines the accuracy and fairness of recommendation \cite{abdollahpouri2020connection,abdollahpouri2019impact}, but also exacerbates the Matthew Effect and the filter bubble through the user-system feedback loop \cite{mansoury2020feedback,gao2023cirs,gao2023alleviating}.

% \begin{figure}[t]
%     \centering
%     \hspace{-8mm}
%     \setlength{\abovecaptionskip}{0.cm}
%     \setlength{\belowcaptionskip}{-1.cm}
%     \includegraphics[width=3.5in]{pics/rec_proportion.pdf}
%     \caption{Illustration of popular bias amplification: We divide items into five groups according to their popularity as recent work \cite{zhang2021causal}, and focus on the most popular group. The chart displays four bars, representing the ratio of the items in the most popular group, the percentage of interactions originating from these items in the training set, and the percentages of these items appearing in recommendations from MF and LightGCN, respectively. }  
%     \label{fg:pro}
% \end{figure}

Given the detrimental impact of popularity bias amplification, a thorough understanding of its root causes is crucial.
Although some recent studies have endeavored to elucidate this, their investigations exhibit significant limitations: 
1)  Some researchers \cite{wei2021model,zhang2021causal,wang2021deconfounded} have investigated popularity bias amplification through causal graphs. However, they merely postulate causal relations between item popularity and model predictions without deeply exploring the underlying mechanisms behind the relations. Moreover, their analyses depend on hypothesized causal graphs, which may be flawed due to the widespread presence of unmeasured confounders \cite{ding2022addressing,zhang2018addressing}. 
2) Other studies \cite{zhou2023adaptive,zhang2023mitigating,Zhang2024General,chen2024graph,chen2024macro} have revealed graph neural network  (GNNs) can exacerbate popularity bias. However, these analyses primarily focus to GNNs rather than the mechanisms of generic recommendation models.

\begin{figure}[t]
  \vspace{-10pt}
  \centering
  \setlength{\abovecaptionskip}{-0.5cm}
  \setlength{\belowcaptionskip}{-0.5cm}
  \includegraphics[width=0.5\textwidth]{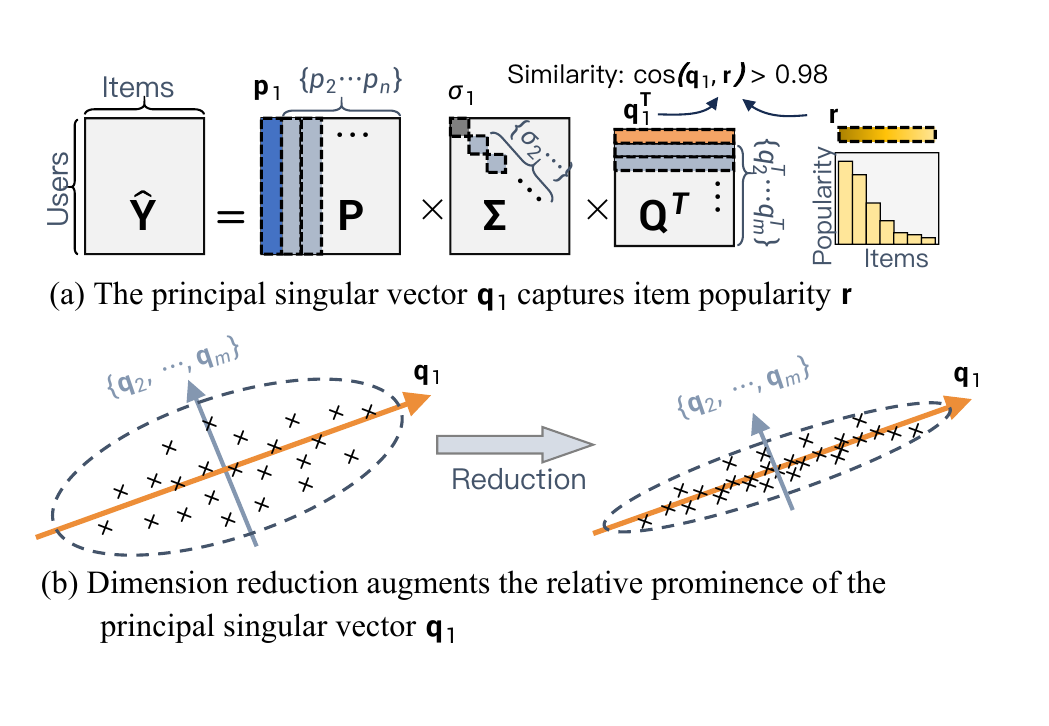}
  \caption{Illustration of two core insights. }  
  \label{fg:main}
  \vspace{-2pt}
\end{figure}

To bridge this research gap, we undertake extensive theoretical and empirical studies on popularity bias amplification. By investigating the spectrum of the ranking score matrix over all users and items predicted by recommendation models, we present the following insights:

\textbf{1) Memorization Effect.} When training a recommendation model on long-tailed data, \textit{the information of item popularity is memorized in the principal spectrum (Figure \ref{fg:main}(a))}. Empirically, we observe that the principal singular vector of the score matrix closely aligns with item popularity, with a cosine similarity consistently exceeding 0.98 across multiple representative recommendation models and datasets. Theoretically, we derive the lower bound of this cosine similarity, demonstrating that the similarity converges to one for highly long-tailed training datasets.

% The phenomenon known as dimension collapse augments the relative prominence of the principal spectrum that captures item popularity, leading to bias amplification (Figure \ref{fg:main}(b)). We reveal that dimension collapse is pervasive in recommendation systems due to two primary reasons: (i) The deliberate low-rank setting of user/item embeddings, employed either to conserve memory or to counteract overfitting, amplifies the impact of the principal spectrum; (ii) The inherent training dynamics of gradient-based optimization prioritize the learning of the principal dimension, while the singular values of other dimensions are easily underestimated. Our further theoretical and empirical analyses establish the relationship between dimension collapse and popularity bias — larger values of principal singular values relative to other singular values result in more serious popularity bias.

\textbf{2) Amplification Effect.} \textit{The phenomenon known as dimension reduction augments the relatively prominence of the principal spectrum that captures item popularity, leading to bias amplification (Figure \ref{fg:main}(b)).} We reveal that dimension reduction is pervasive in RS due to two primary reasons: \romannumeral 1) The deliberate low-rank setting of user/item embeddings, employed either to conserve memory or to counteract overfitting, amplifies the impact of the principal spectrum; \romannumeral 2) The inherent training dynamics of gradient-based optimization prioritize the learning of the principal dimension, while the singular values of other dimensions are easily underestimated. Our further theoretical and empirical analyses establish the relationship between dimension reduction and popularity bias --- larger principal singular values compared to other singular values lead to more popular items on the recommendations.
% \end{itemize}

% \begin{itemize}
%   \item When training a recommendation model on the long-tailed data, \textbf{the information of item popularity is captured in the principal singular vector associated with the largest singular value of the score matrix (Figure \ref{fg:main}(a)).} Empirically, we observe that this singular vector closely aligns with item popularity, with a cosine similarity consistently exceeding 0.98 across multiple recommendation models (\eg MF and LightGCN) and datasets (\eg MovieLens, Douban, and Globo). Theoretically, we derive the lower bound of this cosine similarity, demonstrating that the similarity would converge to one for highly long-tailed datasets.
%   \item  \textbf{The phenomenon known as \textit{dimension collapse} augments the relatively prominence of principle singular vector on prediction, leading to bias amplification (Figure \ref{fg:main}(b)).} Dimension collapse is pervasive in RS due to two primary reasons: 1) The deliberate low-rank setting of user/item embeddings, employed either to conserve memory or to counteract overfitting; and 2) the inherent training dynamics of gradient-based optimization where the learning of the principle dimension is prioritized while the singular values of other dimensions are easily under-estimated. Both empirical and theoretical analyses further ascertain that a pronounced discrepancy between the largest singular value and others intensifies the popularity bias.
% \end{itemize}

Our analysis not only explains the underlying mechanisms of bias amplification but also paves the way for the development of an innovative strategy to counteract this effect. Recognizing that the essence of this amplification lies in the undue contribution of the principal spectrum, we introduce a spectral norm regularizer \cite{yoshida2017spectral} aimed at directly restraining the magnitude of the principal singular value. However, the direct computation of the spectral norm necessitates exhaustive processing of a large score matrix and numerous iterative procedures \cite{yoshida2017spectral, sun2021spectral}, inducing significant computational costs. To address this challenge, we further develop an accelerated strategy by leveraging the intrinsic spectrum properties of the score matrix and matrix transformation techniques. Consequently, our method effectively mitigates popularity bias while imposing limited computational overhead.

In summary, our contributions are:
\vspace{-4pt}
\begin{itemize}
\item Conducting comprehensive analyses to unravel the mechanisms behind popularity bias amplification in recommendation --- item popularity is encoded within the principal singular vector, and its impact is exaggerated due to the dimension reduction phenomenon. 
\item Proposing an efficient method for mitigating the bias amplification through the regulation of the principal singular value. 
\item Performing extensive experiments across seven real-world datasets under three different testing scenarios, demonstrating the superiority of our method in reducing bias and enhancing recommendation quality.
\end{itemize}

\section{Preliminaries}
In this section, we present the background of the recommendation system and popularity bias amplification.

\textbf{Task Formulation.} This work mainly focus on the collaborative filtering (CF) \citep{wu2022survey}, a widely-used recommendation scenario. Consider a RS with a user set $\mathcal U$ and an item set $\mathcal I$. Let $n$ and $m$ denote the total number of users and items. Historical interactions can be expressed by a matrix $\mathbf Y \in\{0,1\}^{n \times m}$, where the element $y_{ui}$ indicates if user $u$ has interacted with item $i$ (\eg click). For convenience, we define the number of interactions of an item as $r_i=\sum_{u\in \mathcal U}{y_{ui}}$, and collect $r_i$ over all items as a popularity vector $\mathbf{r}$. RS targets to suggest items to users based on their potential interests.

\textbf{Recommendation Models.}  Embedding-based models are widely utilized in RS \cite{wu2022survey}. Such models convert user/item attributes (\eg IDs) into $d$-dimensional representations ($\mathbf u_u, \mathbf v_i$), and make predictions using the embedding similarity \cite{wu2022survey}. Given that the inner product is a conventional similarity metric due to its efficiency in retrieval and superior performance \cite{wu2021self, yu2022graph, lin2022improving}, this work also focuses on the inner product for analysis.
% Among the prevalent similarity functions, the inner product \cite{he2016fast} and neural network \cite{he2017neural} stand out. Recent research  \cite{wu2021self,yu2022graph,lin2022improving} indicates that the inner product offers efficient retrieval and often exhibits superior performance. As such, this work focuses on the inner product for analysis. 
Specifically, the model's predicted scores can be formulated as $\hat y_{ui}=\mu(\mathbf u_u^\top \mathbf v_i)$, where $\mu(.)$ denotes an activation function like Sigmoid. $\hat y_{ui}$ represents a user's preference for an item, which is then used for ranking to generate recommendations. For clarity of presentation, we also employ matrix notation. Let matrices $\hat {\mathbf{Y}}$, ${\mathbf{U}}$, $ {\mathbf{V}}$ represent scores over all user-item combinations, embeddings over all users and items, respectively. Model predictions can be succinctly expressed as ${\mathbf{\hat Y}} = \mu({\mathbf{U}}{{\mathbf{V}}^\top})$. 

\textbf{Objective Functions.} Common choices of loss functions for training a recommendation model include point-wise loss such as BCE and MSE \cite{rendle2022item}, and pair-wise loss like BPR \cite{rendle2009bpr}. It is worth noting that BPR can be reconceptualized as a specialized pointwise loss.Concretely, BPR loss is expressed as:
\vspace{-4pt}
\begin{equation}
    {\mathcal{L}_{BPR}} = - \sum\limits_{u \in \mathcal U} {\sum\limits_{i \in \mathcal I,{y_{ui}} = 1} {\sum\limits_{j \in \mathcal I,{y_{uj}} = 0} \log({\mu ({\mathbf{u}}_u^\top{{\mathbf{v}}_i} - {\mathbf{u}}_u^\top{{\mathbf{v}}_j})}) } }
\nonumber
\vspace{-4pt}
\end{equation}
If construct a hyper-item space denoted as $\mathcal I' = \mathcal I \times \mathcal I$ derived from item pairs, and define the embeddings of hyper-items as ${\mathbf{v}'_{ij}} = {{\mathbf{v}}_i} - {{\mathbf{v}}_j}$ and assign new observed interactions to the combinations of users and hyper-items, \ie   ${y'_{u,ij}} = 1$ for $y_{ui}=1\&y_{uj}=0$ and  ${y'_{u,ij}} = 0$ for $y_{ui}=0\&y_{uj}=1$. BPR can be re-written as:
\vspace{-4pt}
\begin{equation}
\vspace{-4pt}
\small
\begin{aligned}
{\mathcal{L}_{BPR}}
    =- \frac{1}{2} \sum\limits_{u \in \mathcal U} \Big(&\sum_{\substack{(i,j)\in \mathcal{I'}\\ y'_{u,ij}=1}} \log(\mu({\mathbf{u}}_u^\top{\mathbf{v}'_{ij}})) 
    +\sum_{\substack{(i,j)\in \mathcal{I'}\\ y'_{u,ij}=0}} \log(\mu(-{\mathbf{u}}_u^\top{\mathbf{v}'_{ij}}))\Big) \nonumber
\end{aligned}
\vspace{-4pt}
\end{equation}
where BPR can be reframed as a specific point-wise loss under the hyper-items space $\mathcal{I'}$. 
Therefore, for convenience, our analyses mainly focus on point-wise loss. But we will also discuss why our proposed debiased method is suitable for BPR (\cf Section \ref{se:disscuss}) and validate its effectiveness in experiments (\cf Section \ref{se:ex}).

\textbf{Popularity Bias Amplification.}  Items' interaction frequency in recommendation data often follows a long-tailed distribution \cite{DBLP:conf/edbt/BorgesS20,ccoba2017visual,steck2011item}. For instance, in a typical Douban dataset, a mere 20\% of the most popular items account for 86.3\% of all interactions. 
 % We discover that a mere 0.6\% of the most popular items, accounting for 20\% of total interactions in data, occupy over 63\% recommendation slots in Douban.
% Readers may refer to Appendix \ref{bias amplify} for more details about such long-tailed distribution.
When models are trained on such skewed data, they tend to absorb and amplify this bias, frequently over-prioritizing popular items in their recommendations. For example, in the Douban dataset using the MF model, 20\% of the most popular items occupy over 99.7\% of the recommendation slots, while a mere 0.6\% of the most popular items occupy more than 63\% (\cf Appendix \ref{appendix: bias amplify} more examples). This notorious effect significantly impacts the recommendation accuracy and fairness, even potentially posing detrimental effects on the entire ecosystem of RS\cite{chen2023bias}. Thus, understanding the underlying mechanisms behind this effect is crucial. 

% Readers may refer to Appendix \ref{bias amplify} for more details about long-tailed distribution and bias amplification effect. 

\section{Understanding Popularity Bias Amplification}
In this section, we conduct thorough analyses to answer:\\
1) How do recommendation models memorize the item popularity? \\
2) Why do recommendation models amplify popularity bias? 

\subsection{Popularity Bias Memorization Effect}

\textit{3.1.1 Empirical Study.} To discern how recommendation models memorize item popularity, we designed the following experiment: 1) We well trained three representative recommendation models, MF \cite{marlin2009collaborative}, LightGCN \cite{he2020lightgcn} and XSimGCL \cite{yu2023xsimgcl}, on three real-world datasets (\cf Section \ref{se:ex} for experimental details); 2) We then performed SVD decomposition on the predicted score matrix, $\hat {\mathbf{Y}}= \mathbf{P}\mathbf{\Sigma} \mathbf{Q}^\top=\sum_{1 \le k \le L} {{\sigma _k}{{\mathbf{p}}_k}{\mathbf{q}}_k^\top}$ where $L=min(n,m)$ and $\sigma_1 \geq \sigma_2 \geq ... \geq \sigma_L$. We further computed cosine similarity between the right principal singular vector $\mathbf{q}_1$ and the item popularity $\mathbf{r}$. The outcomes are showcased in Table \ref{cos_sim}. From these experiments, we draw an impressive observation:

\begin{observation}
  The principal right singular vector $\mathbf{q}_1$ of the matrix $\hat {\mathbf{Y}}$ aligns significantly with the item popularity $\mathbf{r}$. The cosine similarity consistently surpasses 0.98 over multiple recommendation models and datasets.  
  \label{ob1}
\end{observation}

Given the orthogonal nature of different singular vectors, we can deduce that item popularity is almost entirely captured in the principal spectrum. This intriguing phenomenon elucidates how the recommendation model assimilates item popularity from the data and how this popularity influences recommendation outcomes.

% As such, the value of the principle singular value $sigma_1$ plays an important role 

% Given the typically large magnitude of the principal singular value $\sigma_1$, item popularity is bound to exert a significant influence on model predictions. We will elucidate this further in the next subsection.

% \begin{table}[t]
%     \small
%     \centering
%     % \setlength{\abovecaptionskip}{0.1cm}
%     % \setlength{\belowcaptionskip}{-0.2cm}
%     \caption{The cosine similarity between the principal singular vector ($\mathbf{q}_1$) and the item popularity ($\mathbf{r}$) under different backbones and loss functions. BPR loss is not included as the space of hyper-items is too large to be empirically explored.}
%     \begin{tabular}{ccccccc}
%     \toprule 
%     \multirow{2}{*}{\textbf{Backbone}}&\multicolumn{3}{c}{\textbf{MSE}}&\multicolumn{3}{c}{\textbf{BCE}} \\ \cmidrule(r){2-4}\cmidrule(r){5-7} 
%      &\multicolumn{1}{c}{\textbf{Movielens}}&\multicolumn{1}{c}{\textbf{Douban}}&\multicolumn{1}{c}{\textbf{Globo}}&\multicolumn{1}{c}{\textbf{Movielens}}&\multicolumn{1}{c}{\textbf{Douban}}&\multicolumn{1}{c}{\textbf{Globo}}\\
%      \midrule 
%     MF&0.993&0.992&0.993&0.988&0.991&0.989\\
%     LightGCN&0.992&0.990&0.992&0.991&0.988&0.990\\
%     XSimGCL&0.998&0.991&0.990&0.994&0.992&0.985\\
%      %\midrule
%     \bottomrule
%     \label{cos_sim}
%     \end{tabular}
%     \vspace{-20pt} 
% \end{table}
\begin{table}[t]
    \tabcolsep=1.5pt
    \renewcommand\arraystretch{0.95}
    \centering
    \setlength{\abovecaptionskip}{0.1cm}
    \setlength{\belowcaptionskip}{-0.1cm}
    \caption{The cosine similarity between the principal singular vector ($\mathbf{q}_1$) and the item popularity ($\mathbf{r}$) under different backbones and loss functions. }
    \begin{tabular}{cccc|ccc|ccc}
    \toprule 
    \multirow{2}{*}{\textbf{Backbone}}&\multicolumn{3}{c}{\textbf{Movielens}}&\multicolumn{3}{c}{\textbf{Douban}}&\multicolumn{3}{c}{\textbf{Globo}} \\ \cmidrule(r){2-4}\cmidrule(r){5-7}\cmidrule(r){8-10} 
     &\multicolumn{1}{c}{\textbf{MSE}}&\multicolumn{1}{c}{\textbf{BCE}}&\multicolumn{1}{c}{\textbf{BPR}}&\multicolumn{1}{c}{\textbf{MSE}}&\multicolumn{1}{c}{\textbf{BCE}}&\multicolumn{1}{c}{\textbf{BPR}}&\multicolumn{1}{c}{\textbf{MSE}}&\multicolumn{1}{c}{\textbf{BCE}}&\multicolumn{1}{c}{\textbf{BPR}}\\
     \midrule 
    MF&0.993&0.988&0.991&0.992&0.991&0.993&0.993&0.989&0.992\\
    LightGCN&0.992&0.991&0.992&0.990&0.988&0.990&0.992&0.990&0.991\\
    XSimGCL&0.998&0.994&0.995&0.991&0.990&0.992&0.992&0.985&0.989\\
    \bottomrule
    \label{cos_sim}
    \end{tabular}
    \vspace{-26pt} 
\end{table}
\textit{3.1.2 Theoretical Analyses.} Prior to the theoretical validation of observation \ref{ob1}, we posit a power-law hypothesis pertaining to recommendation data: 

\newtheorem{hyo}{Hypothesis}
\begin{hyo}
  The interaction frequency of items in recommendation data follows a power-law distribution (\aka Zipf law) described by $r_g \propto g^{-\alpha}$. 
\end{hyo}
Here $r_g$ signifies the popularity of the $g$-th most popular item, and $\alpha$ is a shape parameter indicating the distribution's slope. Power-law, as a typical long-tailed distribution, is prevalent across various natural and man-made phenomena \cite{clauset2009power}. Recent studies assert that item popularity in RS also aligns with this ubiquitous principle \cite{DBLP:conf/edbt/BorgesS20,ccoba2017visual,steck2011item}. Then we have the following important theorem:
\begin{theorem}[Popularity Memorization Effect]
Given an embedding-based recommendation model with sufficient capacity, when training the model on the data with power-law item popularity, the cosine similarity between item popularity $\mathbf{r}$ and the principal singular vector $\mathbf{q}_1$ of the predicted score matrix is bounded with:
\vspace{-3pt}
\begin{equation}
\small
  \begin{aligned}
    \cos({\mathbf{r}},{{\mathbf{q}}_1}) \ge \frac{{\sigma _1^2}}{{{r_{\max }}\sqrt {\zeta (2\alpha )} }}\sqrt {1 - \frac{{{r_{\max }}(\zeta (\alpha ) - 1)}}{{\sigma _1^2}}} \label{eq:t1}
\end{aligned}
\end{equation}
For $\alpha>2$, this can be further bounded with: 
\vspace{-3pt}
\begin{equation}
\small
  \begin{aligned}
     \cos({\mathbf{r}},{{\mathbf{q}}_1}) \ge \sqrt {\frac{{2 - \zeta (\alpha )}}{{\zeta (2\alpha )}}} \label{eq:t2}
\end{aligned}
\end{equation}
where $r_{\max}$ is the popularity of the most popular item, and $\zeta (\alpha )$ is Riemann zeta function with $\zeta(\alpha) = \sum\limits_{j=1}^{\infty}\frac{1}{j^\alpha}$.   
\label{thm:memory}
\end{theorem}
Proof can be found in Appendix \ref{appendix:thm1}. Notably, as the long-tailed nature of item popularity intensifies (\ie $\alpha \to \infty $ suggesting $\zeta (\alpha ) \to 1$), the right side of Eq. (\ref{eq:t2}) converges to one, implying a near-perfect alignment between $\mathbf{r}$ and $\mathbf q_1$. Even when the data isn't markedly skewed and has a considerable $\zeta (\alpha )$, we typically observe $\sigma_1^2$ to vastly exceed $r_{\max}$, \eg ${\rm{5.6} \times 10^5}$ vs. ${\rm{4.6} \times 10^3}$ in the dataset Movielens (with more examples presented in Appendix \ref{appendix:sigma}). Thus, from Eq. (\ref{eq:t1}), a high similarity between $\mathbf{r}$ and $\mathbf q_1$ emerges. This theorem provides theoretical validation for our observation \ref{ob1}.

%Appendix \ref{appendix:sigma}

\begin{figure*}[t]
    \centering
    % \vspace{-0.5em}
    \setlength{\abovecaptionskip}{-0.03cm}
    \setlength{\belowcaptionskip}{-0.3cm}
    \hspace{-3mm}
    \subfigure[Exploring different dimensions.]{
    \includegraphics[width=2.27in,height=1.45in]{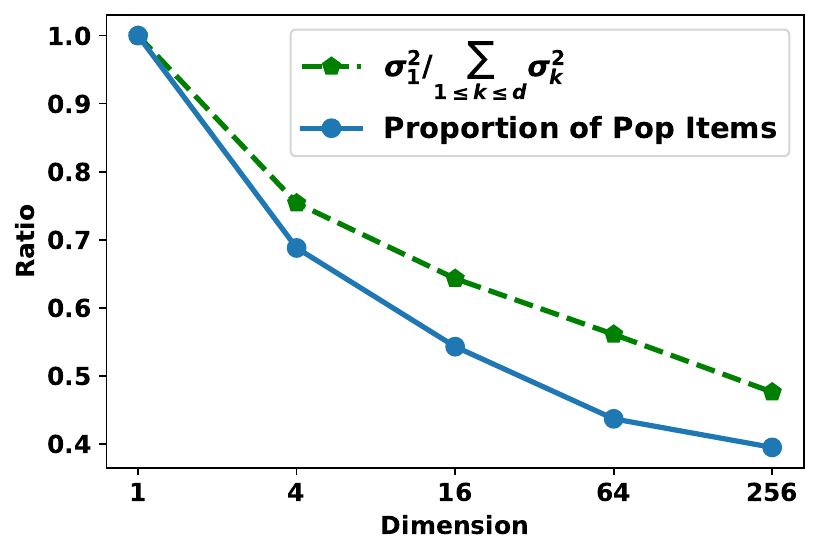}\label{e-a}}
    \subfigure[Exploring different epochs.]{
    \includegraphics[width=2.27in,height=1.45in]{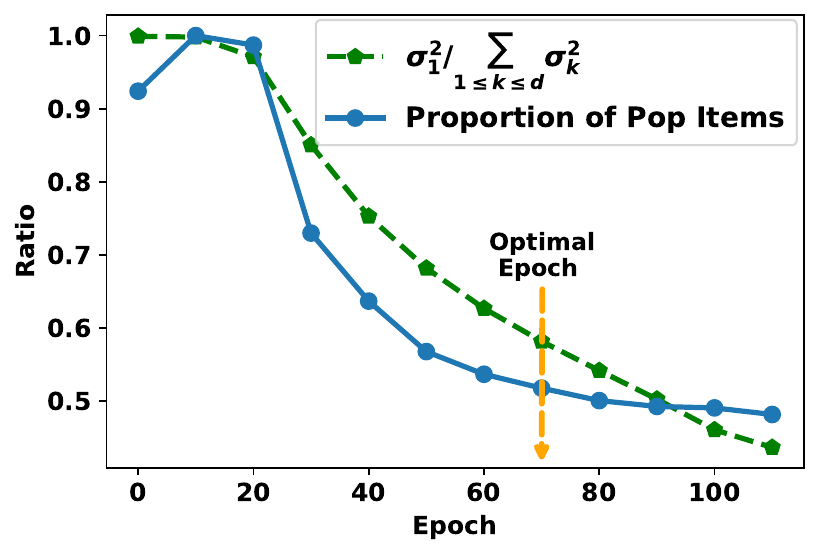}\label{e-b}}
    \subfigure[Exploring singular values.]{
    \includegraphics[width=2.3in,height=1.45in]{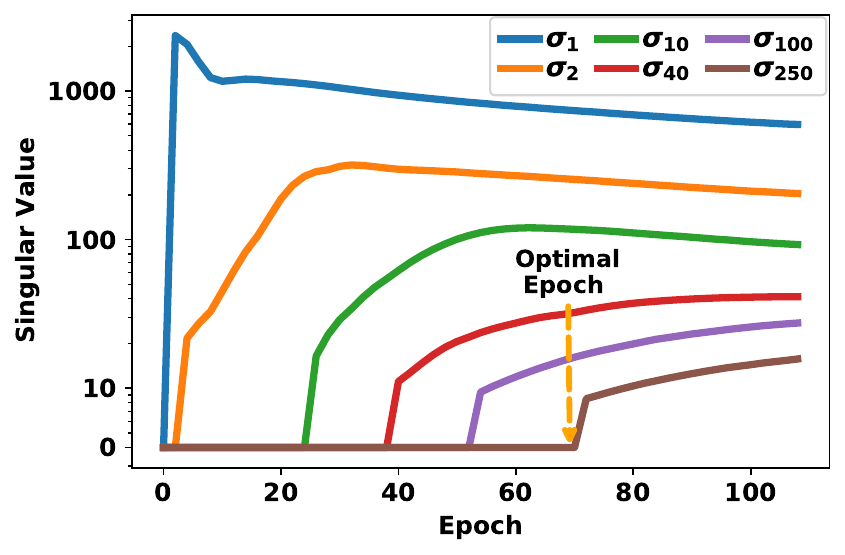}\label{e-c}}
    \caption{Illustration of how dimension reduction impacts popularity bias in Movielens: (a)-(b) the proportion of popular items in recommendations and the ratio of the largest singular value ($\sigma_1^2/\sum_{1\le k\le L} {\sigma_k^2}$) with varying embedding dimensions and training epochs, respectively; (c) how singular values evolves during training.
     }
    \label{a}
\end{figure*}

\subsection{Popularity Bias Amplification Effect}
% Earlier discussions illuminated how item popularity is encoded within the principal singular vector. In this subsection, we further delve into how this singular vector unduly influences predictions through the phenomenon of dimension collapse, ultimately leading to bias amplification. 
% %dimension collapse augments the relative impact of the principal singular vector, leading to bias amplification.
% According to SVD decomposition $\widehat {\mathbf{Y}} = \sum_{1 \le k \le d} {{\sigma _k}{{\mathbf{p}}_k}{\mathbf{q}}_k^\top}$, the dimension collapse diminishes other singular values, thereby augmenting the relative prominence of the principal singular vector. We begin with empirical analyses to show this phenomenon.

Earlier discussions illuminate that the principal spectrum memorizes item popularity. In this subsection, we reveal the phenomenon of dimension reduction in RS, which amplifies the effect of the principal spectrum, leading to popularity bias amplification.

\textit{3.2.1 Empirical Study.} The occurrence of dimension reduction in RS is largely attributable to two factors: 1) explicit low-rank configuration of user/item embeddings \cite{mnih2007probabilistic,he2020lightgcn}, and 2) intrinsic training dynamics associated with gradient-based optimization \cite{chou2023gradient,saxe2019mathematical,arora2019implicit}. Here, we present experiments to validate these points and examine their impacts on popularity bias.

\textbf{Impact of Low-Rank Configuration.} Figure \ref{e-a} displays the proportion of popular items in recommendations from well-trained MF models with varying embedding dimensions $d$. We also present the magnitude of the largest singular value $\sigma_1$ compared with other singular values. We report $\sigma_1^2/\sum_{1 \le k \le L} \sigma_k^2$ as it is easily calculable, where the denominator equals the sum of the diagonal elements of $\hat {\mathbf Y}$. We observe:

\begin{observation} As the model embedding dimension $d$ is reduced, the relative prominence of the principal singular value increases ($\sigma_1^2/\sum_{1 \le k \le L} \sigma_k^2 \uparrow$), and the recommendation increasingly favors popular items.
\end{observation}

This observation reveals the impact of low-rank embeddings. A smaller
$d$ squeezes the dimensions (causing singular values of more dimensions to become zero), thereby relatively amplifying the effect of the principal spectrum. Consequently, item popularity contributes more significantly to ranking, resulting in more severe popularity bias.

% there's a pronounced inclination towards recommending popular items. This can be explained by the relatively increasing influence of $\mathbf q_1$ on predictions, which results in item popularity predominating. It can be seen from the larger value of $\sigma_1^2/\sum_{1 \le k\le d} {\sigma_k^2}$. 

\textbf{Dimension Collapse from Gradient Optimization.} 
Figure \ref{e-c} illustrates the evolution of singular values as training progresses using a gradient-based optimizer; and Figure \ref{e-b} offers a dynamic view of popularity bias and the ratio $\sigma_1^2/\sum_{1 \le k\le L} {\sigma_k^2}$ over the training procedure. We observe:

\begin{observation} The principal singular value grows preferentially and swiftly, while others exhibit a more gradual increment. Notably, many singular values appear to be far from convergence even at the end of the training process. Accordingly, popularity bias is severe at the beginning but exhibits a relative decline as training advances. But even at the end of training, unless an extensive number of epochs are employed (which could result in computational overhead and potential over-fitting), the bias remains pronounced.    
\end{observation}

This phenomenon reveals the dynamic of singular values during gradient optimization --- the principal dimension is prioritized, while singular values of other dimensions are easily under-estimated. This inherent mechanism could readily lead to dimension collapse, relatively enhancing the impact of the principal spectrum, and thereby inducing popularity bias.

% It's evident that the principal singular value grows preferentially and swiftly, while others exhibit a more gradual increment. Notably, many singular values appear to be far from convergence even towards the end of the training process, suggesting an underestimation of these values during model training. Figure \ref{e-b} offers a dynamic view of popularity bias and the ratio $\sigma_1^2/\sum_{1 \le k\le d} {\sigma_k^2}$ over the procedure of training. Popularity bias is severe at the beginning but exhibits a relative decline as training advances. However, even at the end of training, unless an extensive number of epochs are employed (which could result in computational overhead and potential over-fitting), the bias remains pronounced.

\textit{3.2.2 Theoretical Analyses.} 
In this subsection, we focus on establishing a theoretical relationship between singular values and the ratio of popular items in recommendations. For readers interested in the theoretical support of the impact of gradient optimization, we refer them to the Appendix \ref{appendix:gradient}, which are relatively straightforward by invoking recent gradient theory \cite{chou2023gradient,saxe2019mathematical}. For convenience, our analysis here concentrates on the ratio of the most popular item in top-1 recommendations. We have:

\begin{theorem}[Popularity bias amplification]
Given hypothesis 1 and nearly perfect alignment between $\mathbf{q}_1$ and $\mathbf{r}$, the ratio of the most popular item in top-1 recommendations over all users is bounded by: 
\vspace{-3pt}
  \begin{equation}
  \small
    \eta  \ge \frac{1}{n}\phi \left( {\frac{{\sqrt {2\zeta (2\alpha )} }}{{1 - {2^{ - \alpha }}}}(\frac{{\sum\nolimits_{1 \le k \le L} {{\sigma _k}} }}{{{\sigma _1}}} - 1)} \right) \label{eq4}
  \end{equation}
  where $\phi(a)=\sum_{u\in \mathcal U} {\mathbf I[p_{1u}> a]}$ is an inverse cumulative function calculating the number of elements $p_{1u}$ in the left principal singular vector $\mathbf p_{1}$ exceeding a given value $a$, and the function $\mathbf I[.]$ signifies an indicator function. 
\label{thm:amplify}
  \end{theorem}
  The detailed proof is available in Appendix \ref{appendix: thm2}. 
  This theorem vividly showcases the influence of dimension reduction on popularity bias. Essentially, as dimension reduction intensifies the relative prominence of the principle singular value ($\frac{{{\sigma _1}}}{{\sum\nolimits_{1\le k \le L} {{\sigma _k}} }} \uparrow $), the input of the function  $\phi(.)$ decreases ($\frac{{\sqrt {2\zeta (2\alpha )} }}{{1 - {2^{ - \alpha }}}}(\frac{{\sum\nolimits_{1 \le k \le L} {{\sigma _k}} }}{{{\sigma _1}}} - 1) \downarrow $). Given the monotonically decreasing nature of $\phi(.)$, dimension reduction thus escalates the ratio of most popular items in recommendations. Interestingly, the theorem illustrates the impact of a long-tailed distribution on popularity bias. A larger $\alpha$ (indicating a more skewed item popularity distribution) decreases the value of $\frac{{\sqrt {2\zeta (2\alpha )} }}{{1 - {2^{ - \alpha }}}}$, further elevating the lower bound of the ratio, intensifying bias.

\section{Proposed Method}
In this section, we first introduce our proposed debiasing method, followed by a discussion of its properties and a comparison with other debiasing approaches.
  \vspace{-5pt}
  \subsection{ReSN: Regulation with Spectral Norm}
  The above analyses elucidate the essence of the popularity bias amplification --- the undue influence of the principal spectrum. 
  Therefore, the core of an effective debiasing strategy naturally lies on mitigating this effect. To address this, we propose ReSN which leverages \textit{Spectral Norm} Regularizer to penalize the magnitude of principal singular value:
  % one might naturally think of introducing a regularizer that penalizes the magnitude of this singular value. Thus, we propose a novel method ReSN for tackling the bias, incorporating the \textit{spectral norm} as a regularizer:
\vspace{-2pt}
  \begin{equation}
    {\mathcal L_{{\mathop{\rm Re}\nolimits} SN}} = {\mathcal L_R}({\mathbf{Y}},{\mathbf{\hat Y}}) + \beta ||{\mathbf{\hat Y}}||_2^2
\vspace{-2pt}
  \end{equation}
  where ${\mathcal L_R}({\mathbf{Y}},{\mathbf{\hat Y}})$ is original recommendation loss, and $||.||_2$
denote the spectral norm of a matrix measuring its principle singular value; $\beta$ controls the contribution from the regularizer.   

However, there are practical challenges: 1) the $n \times m$ dimensional matrix $\mathbf{\hat Y}$ can become exceptionally large, often comprising billions of entries, making direct calculations computationally untenable; 2) Existing methods to determine the gradient of the spectral norm are iterative \cite{yoshida2017spectral, sun2021spectral}, which further adds computational overhead.

To circumvent these challenges, we make two refinements: 

Firstly, given the alignment of the principal singular vector $\mathbf q_1$ with item popularity $\mathbf r$, the calculating of the spectral norm can be simplified into: $||{\mathbf{\hat Y}}||_2^2 = ||{\mathbf{\hat Y}}{{\mathbf{q}}_1}||^2 \approx ||{\mathbf{\hat Yr}}||^2/||\mathbf r||^2$, where $||.||$ denotes the L2-norm of a vector. It transforms the calculation of the complex spectral norm of a matrix to a simple L2-norm of a vector, avoiding iterative algorithms by leveraging the singular vector property. Further, the item popularity $\mathbf r$ can be quickly computed via ${\mathbf{r}} = {\mathbf{Y}^\top}{{\mathbf{e}}}$, where ${\mathbf{e}}$ represents a $n$-dimension vector filled with ones.

Secondly, we exploit the low-rank nature of the matrix $\mathbf{\hat Y}$. For models based on embeddings,  $\mathbf{\hat Y}$ can be expressed as ${\mathbf{\hat Y}} = \mu ({\mathbf{U}}{{\mathbf{V}}^\top})$, where ${\mathbf{U}}$ and ${{\mathbf{V}}}$ represent the embeddings associated with users and items, respectively, and 
$\mu(.)$ designates an activation function. Our approach turns to penalize the spectral norm of the matrix before the introduction of the activation function. This is motivated by the ease of computation: $||{\mathbf{U}}{{\mathbf{V}}^\top}||_2^2 = ||{\mathbf{U}}({{\mathbf{V}}^\top}{{{\mathbf{\tilde q}}}_1})||^2$, where 
${{{\mathbf{\tilde q}}}_1}$ denotes the right principal vector of the matrix ${\mathbf{U}}{{\mathbf{V}}^\top}$.  By adopting this method, we circumvent the computationally-intensive task of processing the entire matrix $\mathbf{\hat Y}$. Nonetheless, this method introduces a challenge: accurately computing ${{{\mathbf{\tilde q}}}_1}$, since it doesn't inherently align with item popularity. To rectify this, we may simply mirror the calculation of ${\mathbf{q}_1} \leftarrow \frac{{\mathbf{Y}^\top}{{\mathbf{e}}}}{||{\mathbf{Y}^\top}{{\mathbf{e}}}||}$ to ${{{\mathbf{\tilde q}}}_{\mathbf{1}}} \leftarrow \frac{{\mathbf{V}}{{\mathbf{U}}^\top}{{\mathbf{e}}}}{||{\mathbf{V}}{{\mathbf{U}}^\top}{{\mathbf{e}}}||}$.  This approach is clued by our Observation 1 and Theorem 1: a matrix's principal singular vector tends to align with the column sum vector, especially when the vector showcases a long-tailed distribution. 

To empirically validate the accuracy and rationality of the proposed method, we computed the ideal value of $||{\mathbf{U}\mathbf{V}^\top}||_2^2$, as well as the estimated $\frac{||{\mathbf{U}}{{\mathbf{V}}^\top}{\mathbf{V}}{{\mathbf{U}}^\top}{\mathbf{e}}||^2}{{||{\mathbf{U}\mathbf{V}^\top{\mathbf{e}}}||^2}}$ from ReSN, training the MF model with two losses on three datasets. The results are shown in the Table \ref{tab:approx}. According to the table, we found that the actual spectral norms and our approximate estimates are very close across diverse losses and datasets.This indicates that the singular vector $\mathbf{\tilde q}_1$ obtained through $\frac{\mathbf{U}\mathbf{V}^\top\mathbf{e}}{||\mathbf{U}\mathbf{V}^\top\mathbf{e}||}$, serves as an accurate surrogate for the true value of $\mathbf{q}_1$.
Therefore, the estimated regularization term is a accurate surrogate for the spectral norm $||{\mathbf{U}\mathbf{V}^\top}||^2_2$ which validates the precision of this strategy.

In essence, our ReSN optimizes the following loss function: 
\vspace{-6pt}
\begin{equation}
{\mathcal {\tilde L}_{{\mathop{\rm Re}\nolimits} SN}} = {\mathcal L_R}({\mathbf{Y}},{\mathbf{\hat Y}}) + \frac{\beta}{||{\mathbf{V}}{{\mathbf{U}}^\top}{\mathbf{e}}||^2} ||{\mathbf{U}}{{\mathbf{V}}^\top}{\mathbf{V}}{{\mathbf{U}}^\top}{\mathbf{e}}||^2
 \vspace{-10pt}
 \end{equation} 
\subsection{Discussions}
\label{se:disscuss}
The proposed ReSN have the following aspects:

\textbf{Model-Agnostic:} The proposed ReSN is model-agnostic and easy to implement. Given that ReSN introduces merely a regularization term, it can be easily plugged into existing embedding-based methods with minimal code augmentation. 

\textbf{Efficiency:} The regularizer can be fast computed from right to left ---  it predominantly requires the multiplication of a $n \times d$ (or $m \times d$) matrix with a vector. With a time complexity of $O((n+m)d)$, ReSN is highly efficient. Section \ref{ex:efficiency} also provides empirical evidence. The additional time for calculating the regularizer is negligible. 

 \textbf{Suitable for BPR Loss:} As delineated in Section 2, while BPR can be regarded as a specialized point-wise loss, it involves the concept of hyper-items. It means that the regularizer should be conducted on the embedding matrix of hyper-items $V' \in \mathbb R^{m^2\times d}$, \ie $\frac{||{\mathbf{U}}{{\mathbf{V'}}^\top}{\mathbf{V'}}{{\mathbf{U}}^\top}{\mathbf{e}}||^2}{||{\mathbf{V'}}{{\mathbf{U}}^\top}{\mathbf{e}}||^2}$, rather than $\frac{||{\mathbf{U}}{{\mathbf{V}}^\top}{\mathbf{V}}{{\mathbf{U}}^\top}{\mathbf{e}}||^2}{||{\mathbf{V}}{{\mathbf{U}}^\top}{\mathbf{e}}||^2}$. In the following, we will build their approximations. For the numerator part, we have:
 % In fact, we have:
 % As delineated above, BPR can be regarded as a specialized point-wise loss by introducing the concept of hyper-items.  Let's denote the embeddings of these hyper-items with matrix $\mathbf V' \in \mathbb R^{m^2\times d}$, where the ($i\cdot m+j$)-th row corresponds to the embedding of the $ij$-th hyper-item ${\mathbf{v}'_{ij}} = {{\mathbf{v}}_i} - {{\mathbf{v}}_j}$. One might infer that an apt regularizer for the BPR loss would be $\frac{||{\mathbf{U}}{{\mathbf{V'}}^\top}{\mathbf{V'}}{{\mathbf{U}}^\top}{\mathbf{e}}||^2}{||{\mathbf{V'}}{{\mathbf{U}}^\top}{\mathbf{e}}||^2}$, rather than $\frac{||{\mathbf{U}}{{\mathbf{V}}^\top}{\mathbf{V}}{{\mathbf{U}}^\top}{\mathbf{e}}||^2}{||{\mathbf{V}}{{\mathbf{U}}^\top}{\mathbf{e}}||^2}$. 
    \vspace{-3pt}
    \begin{equation}
    \begin{aligned}
  {{{\mathbf{V'}}}^\top}{\mathbf{V'}} &= \sum\limits_{i,j \in I} {({{\mathbf{v}}_i} - {{\mathbf{v}}_j})^\top{{({{\mathbf{v}}_i} - {{\mathbf{v}}_j})}}} = 2m\mathbf V^\top{\mathbf V} - 2m^2{\mathbf{\bar v}}^\top{{{\mathbf{\bar v}}}} \nonumber
  \end{aligned}    
  \vspace{-3pt}
    \end{equation}
    % \\ &= 2m\sum\limits_{i \in I} {{{\mathbf{v}}_i}^\top{\mathbf{v}}_i}  -2(\sum\limits_{i \in I} {{{\mathbf{v}}_i}} )^\top{(\sum\limits_{i \in I} {{{\mathbf{v}}_i}} )}
where $\bar {\mathbf v}=\sum_{i \in I} {{{\mathbf{v}}_i}}/m$ denote the mean vector of the item embeddings. Furthermore, current literature posits that an ideal item representation should emulate a uniform distribution over the unit ball \cite{wang2022towards}. This implies that $\bar {\mathbf v}$ tends to gravitate towards the origin. Thus, ${{{\mathbf{V'}}}^\top}{\mathbf{V'}}$ can be approximated by $\mathbf V^\top{\mathbf V}$ and $||{\mathbf{U}}{{\mathbf{V}'}^\top}{\mathbf{V}'}{{\mathbf{U}}^\top}{\mathbf{e}}||^2$ can be approximated by $||{\mathbf{U}}{{\mathbf{V}}^\top}{\mathbf{V}}{{\mathbf{U}}^\top}{\mathbf{e}}||^2$.

Similarly, for the denominator:
    \vspace{-3pt}
\begin{equation}
    \begin{aligned}
        ||{\mathbf{V'U^\top e}}||^2 &=  \sum\limits_{i,j \in I} {({{\mathbf{v}}_i}{\mathbf{U^\top e}} - {{\mathbf{v}}_j}{\mathbf{U^\top e}})^\top{{({{\mathbf{v}}_i}{\mathbf{U^\top e}} - {{\mathbf{v}}_j}{\mathbf{U^\top e}})}}} \\
        &= 2m \sum\limits_{i \in I}({{\mathbf{v}}_i}{\mathbf{U^\top e}})^\top({{\mathbf{v}}_i}{\mathbf{U^\top e}}) - 2 (\sum\limits_{i\in I} {{\mathbf{v}}_i}{\mathbf{U^\top e}} )^\top(\sum\limits_{i\in I} {{\mathbf{v}}_i}{\mathbf{U^\top e}} )    \\
        &= 2m({\mathbf{VU^\top e}})^\top({\mathbf{VU^\top e}})- 2m^2( {{\mathbf{\bar v}}}{\mathbf{U^\top e}})^\top( {{\mathbf{\bar v}}}{\mathbf{U^\top e}}) \nonumber
    \end{aligned}
    \vspace{-3pt}
\end{equation}
We can deduce ${||{\mathbf{V}'}{{\mathbf{U}}^\top}{\mathbf{e}}||^2}$ can be approximated by ${||{\mathbf{V}}{{\mathbf{U}}^\top}{\mathbf{e}}||^2}$. Consequently, ReSN emerges as a logical regularizer even for the BPR loss. This assertion is also validated by our experiments. 
 
   \begin{table}[t]
    \centering
    \small
    \abovecaptionskip=2pt
    \belowcaptionskip=5pt
    \caption{Comparison between the actual spectral norm and the estimated approximation.}
     \renewcommand\arraystretch{0.95}
    \tabcolsep=2pt
    \begin{tabular}{ccccc}
    \toprule 
     \multirow{2}{*}{\textbf{Datasets}}&\multicolumn{2}{c}{\textbf{MSE}}  &\multicolumn{2}{c}{\textbf{BCE}}\\
    \cmidrule(r){2-3}\cmidrule(r){4-5}&$||{\mathbf{UV^\top}}||_2^2$&$||{\mathbf{U(V^\top{{\mathbf{\tilde q}}_1} )}}||^2$&$||{\mathbf{UV^\top}}||_2^2$&$||{\mathbf{U(V^\top{{\mathbf{\tilde q}}_1} )}}||^2$ \\
     \midrule 
     \textbf{Movielens-1M} &$5.627\times 10^5$ &$5.613 \times 10^5$&$5.629\times 10^5$ &$5.620 \times 10^5$\\
     \textbf{Douban}&$1.160\times 10^7$ & $1.155 \times10^7$ &$1.161\times 10^7$ &$1.157 \times10^7$\\
     \textbf{Globo} &$8.321\times 10^6$ &$8.309\times 10^6$&$8.327\times 10^6$&$8.316\times 10^6$\\
    \bottomrule
    \end{tabular}
    \label{tab:approx}
    \vspace{-10pt}
\end{table}

\textbf{Differences from Methods on Dimensional Collapse:} 
\label{se:com}
% 1) Weight decay, as a widely-used regularizer, penalizes the magnitude of each elements within embeddings or other model parameters. This regularizer typically suppresses all singular values uniformly. In contrast, Our ReSN merely penalizes the largest singular value tailored to mitigate popularity bias. 
Recent studies \cite{chen2023towards,zhang2023mitigating,wang2022towards} has also employed regularizers to alleviate the dimensional collapse of user/item embeddings. Our ReSN diverges from these methods in two key aspects: 1) ReSN imposes constraints directly onto the prediction matrix, unlike the embedding matrix constraints utilized in these methods. This distinction is of significance due to the inherent spectral gap between the embeddings and the prediction matrix. 2) ReSN explicitly modulates the influence of the principal spectrum that captures popularity information, while these methods mainly focuses on promoting embedding uniformity. ReSN directly and solely mitigates the impact of the memorized popularity signal, thus demonstrating high efficacy in mitigating popularity bias; while others may disrupt the spectral structure of the prediction, potentially compromising model accuracy. 

\textbf{Differences from Regularization-based Debiasing methods:} Various regularizers are introduced to combat popularity bias \cite{zhu2021popularity,liu2023popularity,rhee2022countering,zhang2023mitigating}. However, except \cite{zhang2023mitigating} as discussed before, existing approaches are typically heuristic, applying strong constraints to model predictions that may break the model's original spectrum. While it could mitigate popularity bias, this approach may also impair the model's ability to capture other useful signals, significantly compromising recommendation accuracy.
Contrasting this, our ReSN is a light and theoretic-grounding approach --- it motivated by the core reason of bias amplification and only modulates the influence of the principle spectrum.

\begin{table*}[t]
     
    % \begin{adjustbox}{max width=\textwidth}
    \centering
    
    % \hspace{-5mm}
    \setlength{\abovecaptionskip}{0.1cm}
    \setlength{\belowcaptionskip}{0.15cm}
    \tabcolsep=4pt
    \renewcommand\arraystretch{0.9}
    \caption{Performance comparison in terms of NDCG between ReSN and other baselines across seven datasets and three testing paradigms. The ``Com''(refers to ``Common'') represents the paradigm where the training and test datasets are partitioned randomly; ``Deb''(refers to ``Debiased'') represents the paradigm where a debiased test dataset is formulated based on item popularity;  ``Uni''(refers to  ``Uniform-exposure'') represents the paradigm where the test data is uniformly-exposed. The best result is bolded and the runner-up is underlined. The mark `*' denotes the improvement achieved by ReSN over best baseline is significant with $p<0.05$. }
    \begin{tabular}{lcccccccccccc}
    \toprule 
     
    \multirow{2}{*}{}
      & \multicolumn{2}{c}{ \textbf{Movielens}}&\multicolumn{2}{c}{ \textbf{Douban}}  & \multicolumn{2}{c}{ \textbf{Yelp2018}}   &\multicolumn{2}{c}{ \textbf{Gowalla}} &\multicolumn{2}{c}{  \textbf{Globo}}& \textbf{Yahoo}&  \textbf{Coat}  \\ \cmidrule(r){2-3}\cmidrule(r){4-5}\cmidrule(r){6-7}\cmidrule(r){8-9}\cmidrule(r){10-11}\cmidrule(r){12-12}\cmidrule(r){13-13}
& Com& Deb& Com& Deb& Com& Deb& Com& Deb& Com& Deb& Uni& Uni\\
    \midrule 
      MF &0.3572&0.1490&0.0440&0.0116&0.0416&0.0164&0.1182&0.0438&0.1709&0.0028&0.6672&0.5551\\
     Zerosum  &0.3309&0.1411&0.0434&0.0110&0.0415&0.0137&0.1063&0.0421&0.1630&0.0036&0.6665&0.5633\\
     MACR &\underline{0.3732}&0.1647&0.0441&0.0145&0.0404&0.0208&0.1107&0.0545&\textbf{0.1782}&\underline{0.0253}&0.6714&0.5661\\     PDA &0.3688&\underline{0.1662}&0.0446&0.0171&\underline{0.0437}&\underline{0.0229}&0.1283&\underline{0.0675}&\underline{0.1725}&0.0243&\underline{0.6756}&0.5676\\
 InvCF &0.3723&0.1567&\underline{0.0450}&0.0152&0.0433&0.0183&0.1302&0.0592&{0.1671}&0.0194&0.6519&\underline{0.5715}\\
 IPL&0.3618&0.1621&0.0442&\underline{0.0173}&0.0419&0.0219&\underline{0.1318}&0.0623&0.1715&0.0203&0.6691&0.5602\\
 ReSN&\textbf{0.3857*}&\textbf{0.1745*}&\textbf{0.0456*}&\textbf{0.0186*}&\textbf{0.0445*}&\textbf{0.0254*}&\textbf{0.1343*}&\textbf{0.0703*}&0.1682&\textbf{0.0256*}&\textbf{0.6792*}&\textbf{0.5871*}\\
    \bottomrule
    \end{tabular}\label{performance}
    % \end{adjustbox}
    \end{table*}

\section{Experiments}\label{se:ex}
We conduct experiments to address the following questions: \\
 \textbf{RQ1}: How does ReSN perform compared with other methods?\\
 \textbf{RQ2}: Is ReSN suitable for diversified loss functions and backbones?\\
 \textbf{RQ3}: What is the impact of regularizer coefficient $\beta$?\\
\textbf{RQ4}: How is the efficiency of ReSN ?
\vspace{-5pt}
\subsection{Experiment Settings}
\textbf{Datasets and Metrics.} We adopt seven real-world datasets, Yelp2018 \cite{he2020lightgcn}, Douban \cite{song2019session}, Movielens \cite{yu2020graph}, Gowalla \cite{he2017neural_a}, Globo \cite{de2018news}, Yahoo!R3 \cite{marlin2009collaborative} and Coat \cite{schnabel2016recommendations} for evaluating our model performance. Details about these datasets refer to Appendix \ref{appendix:dataset}. 

We adopt three representative testing paradigms for comprehensive evaluations: 1) \textbf{Common}: We employ the conventional testing paradigm in RS, wherein the datasets are randomly partitioned into training (70\%), validation (10\%), and testing (20\%). We also report the accuracy-fairness trade-off in this setting. 2) \textbf{Debiased}: Closely referring to \cite{wei2021model,bonner2018causal,zheng2021disentangling}, we sample an debiased test set where items are uniformly distributed, aiming to evaluate the model's efficacy in mitigating popularity bias. 3) \textbf{Uniform-exposure}: We also adopt the uniform exposure paradigm for model testing as the recent work \cite{zhang2023invariant,wang2024Distributionally,lin2024recrec}. Notably, the datasets Yahoo!R3 and Coat contain a small dataset collected through a random recommendation policy. Such data isolate the popularity bias from uneven exposure, offering a more precise estimation of user preferences. Consequently, we train our recommendation model on conventionally biased data and then test it on these uniformly-exposed data.

For evaluation metrics, we adopt the widely-used \textbf{NDCG@K} for evaluating accuracy \cite{krichene2020sampled}. We simply adopt $K=5$ for Yahoo and Coat datasets and $K=20$ for the other datasets as recent work \cite{he2020lightgcn,zhang2023invariant,yu2023xsimgcl}. We observe similar results with other metrics. We also employ the \textbf{ratio of pop/unpopular items} for illustrating the severity of popularity bias in recommendations. Here we closely refer to recent work \cite{zhang2021causal} to define popular and unpopular items. 
% divide items into several groups, and simply consider the items in the first group as popular items. Specifically, we 
We sort the items according to their popularity in descending order, and divide items into five groups ensuring the aggregated popularity of items within each group is the same. We define the items in the most popular groups as popular items, while the others as unpopular.

\textbf{Baselines.} The following methods are compared: 1) \textbf{MACR} (KDD'21 \cite{wei2021model}), \textbf{PDA} (SIGIR'21 \cite{zhang2021causal}): the representative causality-based debiasing methods, which posit a causal graph \cite{pearl2009causality} for the recommendation procedure and leverage causal inference to mitigate popularity bias accordingly; 2) \textbf{InvCF} (WWW'23  \cite{zhang2023invariant}): the SOTA method that addresses popularity bias by disentangling the popularity from user preference. 3) \textbf{Zerosum} (Recsys'22 \cite{rhee2022countering}) , \textbf{IPL} (SIGIR'23 \cite{liu2023popularity}): the representative methods based on regularizers, which penalize the score differences or constrain the ratio of the predicted preference with the exposure. 
    % \item \textbf{}: a regularization-based method, which regularizes the proposed IPL criterion by inverse-propensity-score-based estimation.

\begin{table}[t]
\setlength{\tabcolsep}{1.2pt}
    \renewcommand\arraystretch{0.9}
\setlength{\abovecaptionskip}{0.1cm}
\setlength{\belowcaptionskip}{-0.2cm}
\caption{NDCG@20 comparison with methods for addressing Dimension Collapse under the debiased testing paradigm.}
    \begin{tabular}{lccc}
    \toprule
&\textbf{Movielens}&\textbf{Douban}&\textbf{Gowalla}\\
    \midrule
    MF&0.1529&0.0116&0.0438\\
    nCL&0.1572&0.0112&0.0451\\
DirectAU&\underline{0.1691}&\underline{0.0131}&\underline{0.0622}\\
ReSN&\textbf{0.1788}&\textbf{0.0188}&\textbf{0.0712}\\
    \bottomrule
    \end{tabular}\label{performance_DC}
\end{table}

For fair comparisons, we implement all compared methods with uniform MF backbone and MSE loss. We also explore the performance with other backbones and losses in subsection \ref{se:ex2}. Besides above baselines, we also compare our method with the methods on mitigating dimension collapse, including nCL \cite{chen2023towards} and DirectAU \cite{wang2022towards}; and the debiasing methods tailored for GNN-based methods including APDA \cite{zhou2023adaptive} and GCF\textsubscript{logdet} \cite{zhang2023mitigating} when using GNN-based backbones.

 \textbf{Parameter Settings.} The embedding dimension $d$ is 256 while other dimensions are explored in \ref{appendix:dimension_performance} . Grid search is utilized find the optimal hyperparameters. 
More details refer to Appendix \ref{appendix:parameter}
\subsection{Performance Comparison (RQ1)}
%In this section, we conduct multiple experiments to validate the superiority of ReSN over baselines.
\textbf{Comparison under three testing paradigms.} Table \ref{performance} showcases the NDCG@20 comparison across seven datasets over three testing paradigms. Under the Common testing paradigm, our ReSN, with few exceptions, consistently outperforms compared methods. This superior performance can be attributed to the rigorous theoretical foundations of ReSN, which pinpoint and address the root cause of bias amplification. By curbing this bias amplification, ReSN achieves significant improvements in recommendation accuracy. Transitioning to the Debiased and Uniform-exposure testing paradigms, the improvements by ReSN become even more impressive, demonstrating its effectiveness in mitigating popularity bias. 

\begin{figure}[t]
\abovecaptionskip=-0pt
\belowcaptionskip=-3pt
\includegraphics[width=0.49\textwidth]{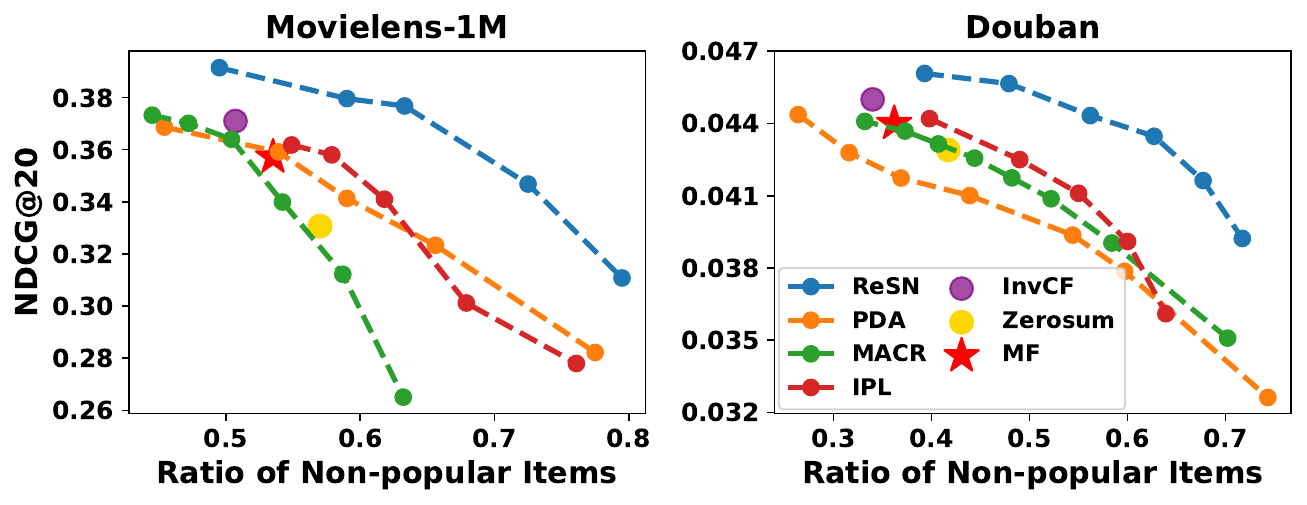}
\caption{Pareto curves of compared methods illustrating the trade-off between accuracy and fairness under the common testing paradigm.}
\label{fig:pareto}
\vspace{-15pt}
\end{figure}

\textbf{Exploring Accuracy-fairness Trade-off.} Given the conventional accuracy-fairness trade-off observed in RS, we delve deeper into examining this effect across various methods. After well-training various methods with differing hyper-parameters (details of hyper-parameters tuning refer to Appendix \ref{appendix:parameter}), we depict the Pareto frontier in Figure \ref{fig:pareto}. It highlights the relationship between accuracy (NDCG@20) and fairness (ratio of unpopular items) under the Common testing paradigm. Here, positions in the top-right corner indicate superior performance.
We observe that ReSN exhibits a more favorable Pareto curve in comparison to other baselines. When fairness is held constant, ReSN showcases superior accuracy. Conversely, when accuracy is fixed, ReSN delivers enhanced fairness. This suggests that ReSN effectively navigates the fairness-accuracy trade-off, primarily through its capability to counteract popularity bias amplification --- it only mitigates the effect of the principle spectrum without disturbing other spectrum. 

\textbf{Compared with the Methods on Tackling Dimension Collapse.} Table \ref{performance_DC} shows the results of our ReSN compared with existing methods on tackling dimension collapse on debiased testing paradigm. nCL and DirectAU can indeed mitigate the popularity bias. However, their performance is inferior to ReSN. The reason is that our ReSN is designed for debiasing, directly modulating the effect of the item popularity on predictions, and thus yielding better performance. 

\subsection{Adaptability Exploration (RQ2)}
\label{se:ex2}
\begin{table}[t]
    \centering
    % \small

    \setlength{\tabcolsep}{1.5pt} 
        \renewcommand\arraystretch{0.9}
\setlength{\abovecaptionskip}{0.1cm}
\setlength{\belowcaptionskip}{-0.2cm}
    \caption{NDCG@20 comparison with GNN-based backbones (LightGCN, XSimGCL) under the debiased testing paradigm.}
    \begin{tabular}{lcccccc}
    \toprule
    &\multicolumn{2}{c}{\textbf{Movielens}} &\multicolumn{2}{c}{\textbf{Douban}} &\multicolumn{2}{c}{\textbf{Gowalla}}\\\cmidrule(r){2-3}\cmidrule(r){4-5}\cmidrule(r){6-7}
    & {{LGCN}}&{{XSGCL}} & {{LGCN}}&{{XSGCL}} & {{LGCN}}&{{XSGCL}}\\
    
    \midrule 
    Backbone&0.1531&0.1686&0.0117&0.0132&0.0446&0.0563\\
    Zerosum&0.1363&0.1438&0.0112&0.0129&0.0437&0.0498\\
    MACR&0.1682&0.1692&0.0157&0.0164&0.0543&0.0623\\
    PDA&\underline{0.1684}&\underline{0.1732}&\underline{0.0182}&0.0190&\underline{0.0689}&\underline{0.0732}\\ 
    InvCF&0.1602&0.1672&0.0153&0.0169&0.0599&0.0687\\
    IPL&0.1653&0.1701&0.0166&\underline{0.0193}&0.0642&0.0699\\
    APDA&0.1657&0.1713&0.0156&0.0189&0.0468&0.0522\\
    GCF\textsubscript{logdet}&0.1672&0.1724&0.0124&0.0141&0.0403&0.0492\\
    
ReSN&\textbf{0.1758}&\textbf{0.1810}&\textbf{0.0194}&\textbf{0.0202}&\textbf{0.0717}&\textbf{0.0763}\\
    \bottomrule
    \end{tabular}\label{performance_LGN}
    \vspace{-0pt}
\end{table}

\begin{table}[t]
    \renewcommand\arraystretch{0.9}
\setlength{\abovecaptionskip}{0.1cm}
\setlength{\belowcaptionskip}{-0.2cm}
    \caption{NDCG@20 comparison with different Loss functions under  the debiased testing paradigm.}
    \tabcolsep=2pt
    \begin{tabular}{lcccccc}
    \toprule
    &\multicolumn{2}{c}{\textbf{Movielens}} &\multicolumn{2}{c}{\textbf{Douban}} &\multicolumn{2}{c}{\textbf{Gowalla}}\\ \cmidrule(r){2-3}\cmidrule(r){4-5}\cmidrule(r){6-7}&{+BCE}&+BPR&+BCE&+BPR&+BCE&+BPR \\
    \midrule 
    MF&0.1529&0.1540&0.0117&0.0120&0.0432&0.0431\\
    Zerosum&0.1472&0.1498&0.0109&0.0106&0.0423&0.0425\\
    {MACR}&\underline{0.1682}&0.1629&0.0155&0.0149&0.0574&0.0546\\
{PDA}&0.1635&\underline{0.1633}&\underline{0.0176}&0.0173&\underline{0.0661}&\underline{0.0675}\\
    {InvCF}&0.1574&0.1582&0.0153&0.0154&0.0553&0.0583\\
    {IPL}&0.1612&0.1628&0.0173&\underline{0.0177}&0.0612&0.0626\\{ReSN}&\textbf{0.1788}&\textbf{0.1693}&\textbf{0.0188}&\textbf{0.0180}&\textbf{0.0712}&\textbf{0.0702}\\
    \bottomrule    
    \end{tabular}\label{performance_loss}
    \vspace{-5pt}
    \end{table}
    % \hspace{7pt}   

\begin{figure}[t]
\vspace{-10pt}
\setlength{\abovecaptionskip}{-0.1cm}
\setlength{\belowcaptionskip}{-0.1cm}
\includegraphics[width=0.5\textwidth,height=0.44\linewidth]{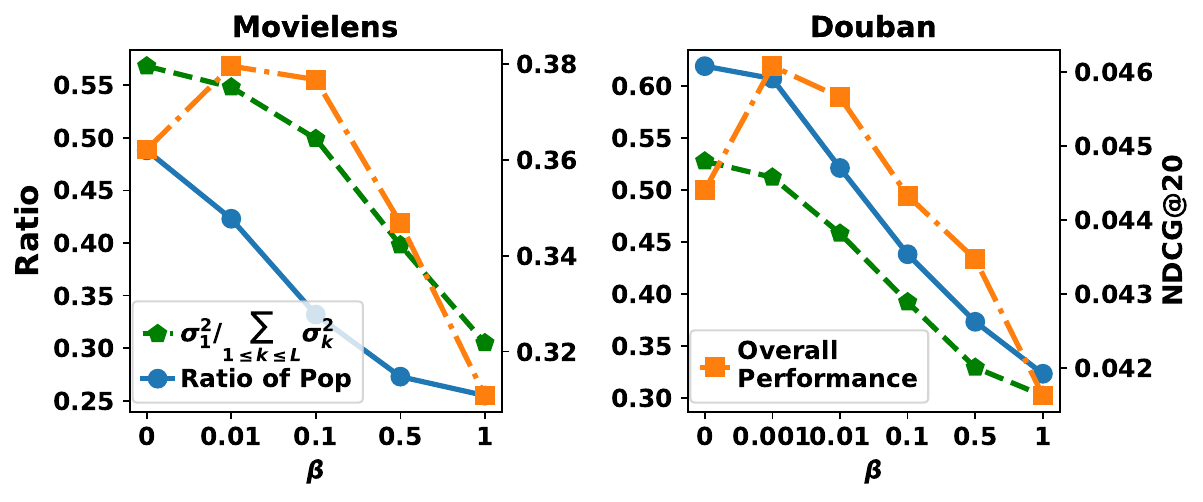}
\caption{The proportion of popular items in recommendations and the ratio of the largest singular value ($\sigma_1^2/\sum_{1\le k\le L} {\sigma_k^2}$) and NDCG@20 with varying $\beta$.}
\label{performance_beta2}

\end{figure}
\begin{table}[t]
    \centering
\setlength{\abovecaptionskip}{0.05cm}
\setlength{\belowcaptionskip}{-0.1cm}
    \renewcommand\arraystretch{0.9}
    \caption{Running time comparison (s/Epoch), where ReSN-Direct directly calculates spectral norm without acceleration.}
    \begin{tabular}{lccc}
    \toprule
&\textbf{Movielens}&\textbf{Douban}&\textbf{Gowalla}\\
    \midrule
    MF&0.177&2.098&0.634\\
    ReSN&0.181&2.124&0.675\\
ReSN-Direct&649&59239&15280\\
\midrule
Speedup Ratio&3585&27890&22637\\
 \bottomrule
    \end{tabular}\label{tab:time_compare}
    \vspace{-9pt}
\end{table}
To investigate the adaptability of ReSN, we evaluate it with various backbones and loss functions under debiased testing paradigm. Table \ref{performance_LGN} showcases the performance of ReSN under LightGCN \cite{he2020lightgcn} and XSimGCL \cite{yu2023xsimgcl}, where APDA \cite{zhou2023adaptive} and GCF\textsubscript{logdet} \cite{zhang2023mitigating} that tailored for GNN-based backbones are included. Besides, Table \ref{performance_loss} depicts the results with BCE and BPR losses. Notably, ReSN consistently outperforms compared methods, irrespective of the chosen backbone or loss function. These findings affirm the great adaptability of ReSN, underscoring its ability to seamlessly integrate with diverse recommendation models. Our ReSN also outperforms those debiasing methods tailored for GNN-based methods. They reason is that they only consider to mitigate bias amplification raised by GNNs while ignoring the bias from the generic recommendation mechanism. 

%Moreover, these empirical outcomes validate our theoretical insights that ReSN is compatible with the BPR.

\subsection{Hyperparameter Study (RQ3)}
\label{se:hyper}
Figures \ref{performance_beta2} presents the recommendation accuracy, the ratio of popular items, and the ratio of principle singular value ($\sigma_1^2/\sum_{1 \le k\le L} {\sigma_k^2}$) in ReSN as the hyperparameter $\beta$ varies. Notably, as $\beta$ increases, the ratio of principle singular value and the severity of popularity bias reduces. This trend affirms the efficacy of our regularizer. Regarding recommendation accuracy, it initially rises and then declines with an increase in $\beta$. This can be attributed to the fact that popularity bias isn't intrinsically detrimental \cite{zhang2021causal,zhao2022popularity}. Indeed, item popularity can also convey beneficial information about the item's appeal or quality, which ought to be retained. Hence, the strategic approach for popularity bias lies in mitigating its bias amplification, rather than eliminating it entirely. That is also our target. 
\vspace{-4pt}
\subsection{Efficiency Study (RQ4)}
\label{ex:efficiency}
To further validate the effectiveness of our acceleration strategy, we test the running time per epoch of ReSN and the original brute-force strategy employed to compute the gradient of the spectral norm. Also, we present the baseline MF for comparison. The results are presented in Table \ref{tab:time_compare}. As can be seen, our acceleration strategy achieves over 3600, 27000 and 22000 times impressive speed-up in both datasets, respectively. Moreover, compared with MF, our ReSN does not incur much computational overhead.

\section{Related Work}

\textbf{Analyses on Popularity Bias}. In RS, items frequently exhibit a long-tailed distribution in terms of interaction frequency. Models trained on skewed data are susceptible to inheriting and exacerbating such bias \cite{jannach2015recommenders,zhu2021popularity,abdollahpouri2019impact,abdollahpouri2020connection,zhu2022evolution,wu2024effectiveness}.  
% This phenomenon has been empirically studied and verified by numerous researchers.  Some studies, such as \cite{jannach2015recommenders,zhu2021popularity}, investigate the influence of popularity bias across different recommendation models. Others, like \cite{abdollahpouri2019impact,abdollahpouri2020connection}, delve into the relationship between popularity bias and fairness. \citet{zhu2022evolution} examine the evolving nature of popularity bias during the feedback loop of RS. 
 % \citet{wang2021deconfounded} used yet another causal graph to highlight bias amplification, focusing on the impact of users' historical long-tailed distribution across item groups.
The crux of tackling popularity bias lies in understanding why and how recommendation models intensify popularity bias. Several recent efforts aim to elucidate this. Among these, causality-based investigations stand out. For instance, \citet{zhang2021causal,wang2021deconfounded} developed a causal graph of the data generative process, attributing the amplification of popularity bias to a confounding effect; \citet{wei2021model} presented an alternate causal graph, exploring the direct and indirect causal influence of popularity bias on predictions. A common limitation among these causality-based methods is their surface-level engagement with the causal relationships among variables, rather than delving deeper into the underlying mechanisms. For example, these studies usually operate on the assumption that item popularity directly affects predictions. However, the specifics of how and why predictions memorize and are influenced by item popularity remain largely unexplored. Worse still, their effectiveness hinges on the accuracy of their respective causal graphs, which might not always hold due to the unmeasured confounders \cite{ding2022addressing,li2023balancing}.

There were other investigations into popularity bias. For instance, \citet{zhu2021popularity} demonstrate that model predictions inherit item popularity, yet they failed to elucidate the amplification. Also, their conclusions rely on a strong assumption that the preference scores maintain same distribution across different user-item pairs. The study by \cite{ohsaka2023curse} shed light on the limited expressiveness of low-rank embeddings, giving clues of popularity bias in recommendations. Yet they did not factor in the impact of long-tailed training data. In fact, popularity bias origins from long-tailed data \cite{zhang2021causal,zhu2021popularity}, amplified during training, which would be more serious than the theoretically analyses presented in \cite{ohsaka2023curse}. Some efforts \cite{chen2023adap,kim2023test} examined popularity bias through embedding magnitude, their theoretical analysis can only applied in the early stages of training.  Other researchers delved into how graph neural networks amplify popularity bias through influence functions \cite{chen2023graph}, the hub effect \cite{zhou2023adaptive} or dimensional collapse \cite{zhang2023mitigating}. However, their conclusions can not be extended to general recommendation models. 

\textbf{Methods on Tackling Popularity Bias.} Recent efforts on addressing popularity bias are mainly four types: 1) Causality-driven methods assume a causal graph to identify popularity bias and employ causal inference techniques for rectification. While they have demonstrated efficacy, their success is closely tied to the accuracy of the causal graph. This poses challenges due to the prevalence of unmeasured confounders \cite{ding2022addressing,li2023balancing,ning2024debiasing}.
 2) Propensity-based methods \cite{schnabel2016recommendations,gruson2019offline,chen2021autodebias,zhang2023model,Wang2024Causally} adjust the data distribution by reweighting the training data instances. While this approach directly negates popularity bias in the data, it may inadvertently obscure other valuable signals, such as item quality. Consequently, these methods often underperform compared to causality-driven ones. 
 3) Regularizer-based methods \cite{abdollahpouri2017controlling,zhu2021popularity,rhee2022countering,liu2023popularity,Jin2024Inforank} constrain predictions by introducing regularization terms. For example, \citet{zhu2021popularity} employs a Pearson coefficient regularizer to diminish the correlation between item popularity and model predictions; \citet{zhang2023mitigating} adopts a regularizer for mitigating embeddings collapse; \citet{rhee2022countering} proposes to regularize the score differences; \cite{liu2023popularity} constrains the predictions with IPL criterion. As discussed in section \ref{se:com}, their constraints are too strong, may significantly compromising accuracy.
 % A potential pitfall of them is the imposition of overly stringent constraints on predictions, which might heavily compromise the accuracy. 
 4) Disentanglement-based methods \cite{zhang2023invariant,xv2022neutralizing,chen2022co} target at learning disentangled embeddings that segregate the influence of popularity from genuine user preferences.  While promising, achieving a perfect disentanglement of popularity bias from true preferences remains a formidable challenge in RS. 

 Among the related work, the one most closely related to ours is \cite{zhang2023mitigating}, but we emphasize that our work differs in two key aspects: 1) Their theoretical justification of bias amplification focuses solely on GNNs, whereas our analysis applies to generic recommendation mechanisms. 2) Their regularizer aims to mitigate collapse of user/item embeddings, while our ReSN specifically targets the mitigation of the principal spectrum's influence. Section \ref{se:disscuss} provides a detailed discussion of these differences, demonstrating that ReSN is more effective in debiasing. Table \ref{performance_DC} also offers empirical evidence supporting our claims.
\section{Conclusion}
In this study, we delve into the root cause of popularity bias amplification. Our analyses offer two core insights: 1) Item popularity is encoded in the principal spectrum of model predictions; 2) The phenomenon of dimension reduction accentuates the influence of the principal spectrum. Based on these insights, we introduce ReSN, an efficient technique aimed at mitigating popularity bias by penalizing the principle singular value. A potential limitation of our study pertains to the static perspective on popularity bias, neglecting its dynamic nature as it evolves temporally.  It could be more insightful to investigate the mechanism of bias amplification in the context of temporal sequential recommendations, and to examine its evolution during the feedback loop.
\begin{acks}
This work is supported by the National Natural Science Foundation of China (62372399, 62476244) and the advanced computing resources provided by the Supercomputing Center of Hangzhou City University.
\end{acks}

\bibliographystyle{ACM-Reference-Format}

\bibliography{sample-base}

\appendix
\newpage
\section{Theoretical Analysis}\label{appendix:theory}

\subsection{Proof of Theorem 1}
\label{appendix:thm1}
The proof procedure of Theorem 1 consists of four parts: 

1) we first showcase the relations between $\mathbf Y$ with $\mathbf {\hat Y}$ under the condition in Theorem 1 and transform $cos(\mathbf r, \mathbf  q_1)$ into $cos(\mathbf r, {\dot {\mathbf q}}_1) $; 

2) We then derive the preliminary lower bound of the $cos(\mathbf r,  {\mathbf  q}_1)$ as $ cos(\mathbf r,  {\mathbf q}_1) \ge \frac{{{  \sigma _1}{{\mathbf{e}}^\top}{{\mathbf{  p}}_1}}}{{{r_{\max }}\sqrt {\zeta (2\alpha )} }}$; 

3) We further utilize the property of $\mathbf {Y}$ to give the lower bound of ${{\mathbf{e}}^\top}{{\mathbf{  p}}_1}$ as
${{\mathbf{e}}^\top}{{\mathbf{  p}}_1} \ge {  \sigma _1}\sqrt {1 - \frac{{{r_{\max }}(\zeta (\alpha ) - 1)}}{{  \sigma _1^2}}}$;

4) Finally, we demonstrate $  \sigma _1^2 \ge {r_{\max }}$, and give $\cos ({\mathbf{r}},{{\mathbf{ q}}_1}) =\cos ({\mathbf{r}},{{\mathbf{ q}}_1}) \ge \sqrt {\frac{{2 - \zeta (\alpha )}}{{\zeta (2\alpha )}}}$ when $\alpha >2$. 

\textbf{Part 1:}  \textbf{the spectral relation between $\mathbf Y$ and $\mathbf {\hat Y}$}. Here we focus on an embedding-based model with sufficient capacity and optimize it with MSE loss \footnote{In practice, we may introduce weight decay or negative sampling for acceleration or mitigating overfitting. But they are not our focus here we simply take the original loss for theoretical analyses.}: ${L_R} = ||{\mathbf{Y}} - {\mathbf{\hat Y}}||_F^2$. For other losses, like BCE loss,which can be approximated to MSE Loss via Taylor expansion \cite{linnainmaa1976taylor}. Based on the theorem of PCA (Principal Component Analysis) \cite{abdi2010principal}, the optimal ${{\mathbf{\hat Y}}}$ will have the same spectrum as the principal dimensions of ${\mathbf{Y}}$. That is, their principal dimensional singular value and vector match up. 
Specifically. ${\mathbf{Y}}$ can be written as:
\begin{equation}
{\mathbf{Y}} = {\dot{\mathbf P}{ \dot \Sigma }}{{\dot{\mathbf Q}}^{\top} = \sum\limits_{k =1}^{ L}  {{\dot \sigma _k}{{\dot{\mathbf p}}_k}{\dot{\mathbf q}}_k^\top}}
\end{equation}
through SVD decomposition.

Because of this relation, when analyzing the principle spectral property of ${{\mathbf{\hat Y}}}$, we can instead shift our focus and look at ${\mathbf{Y}}$, making our job easier.

\textbf{Part 2: preliminary bound of $cos(\mathbf r, \mathbf q_1)$}. Note that $\mathbf r$ can be written as ${\mathbf{r}} = {{\mathbf{Y}}^\top}{\mathbf{e}}$, where $\mathbf{e}$ denotes the $n$-dimension vector filled with ones. We have:
\begin{equation}
 \cos ({\mathbf{r}},{{\dot{\mathbf q}}_{\mathbf{1}}}) = \frac{{{{\mathbf{e}}^\top}{\mathbf{Y}}{{\dot{\mathbf q}}_{\mathbf{1}}}}}{{||{\mathbf{r}}||}} = \frac{{{{\mathbf{e}}^\top}\sum\limits_{k =1}^{ L} {{\dot \sigma _k}{{\dot{\mathbf p}}_k}{\dot{\mathbf q}}_k^\top} {{\dot{\mathbf q}}_{\mathbf{1}}}}}{{||{\mathbf{r}}||}} = \frac{{{\dot \sigma _1}{{\mathbf{e}}^\top}{\dot{\mathbf p}_1}}}{{||{\mathbf{r}}||}}
\end{equation}
where the last equation holds due to the orthogonal of different singular vectors. Further, considering the power-law nature of item popularity, we have:
\begin{equation}
||{\mathbf{r}}|| = \sqrt {\sum\limits_{i =1}^ m {r_i^2} }  = r_{\max}\sqrt {\sum\limits_{i =1}^ m {{i^{ - 2\alpha }}} }  \le r_{\max}\sqrt {\zeta (2\alpha )}
\end{equation}
By combining the above two formulas, we have: 
\begin{equation}
  cos(\mathbf r, { \mathbf q}_1) = cos(\mathbf r,\dot{ \mathbf q}_1) \ge \frac{{{\sigma _1}{{\mathbf{e}}^\top}{{\dot{\mathbf p}}_1}}}{{{r_{\max }}\sqrt {\zeta (2\alpha )} }} \label{eq:p2}
\end{equation}

\textbf{Part 3: bound of ${{\mathbf{e}}^\top}{{{\mathbf p}}_1}$}. We first demonstrate that for the matrix ${\mathbf{Y}}$, we can always find a non-negative principal singular vector ${{{\dot{\mathbf q}}_{{1}}}}$. Let define matrix ${\mathbf{Z}} = {{\mathbf{Y}}^\top}{\mathbf{Y}}$. We can find each element in ${{z_{kl}}}$ is non-negative. Note that:
\begin{equation}
\dot \sigma _1^2 = \mathop {\max }\limits_{||{\dot{\mathbf q}}|| = 1} {{\dot{\mathbf q}}^\top}{{\mathbf{Y}}^\top}{\mathbf{Y}\dot{\mathbf q}} = \mathop {\max }\limits_{||{\dot{\mathbf q}}|| = 1} \sum\limits_{k=1}^m \sum\limits_{l=1}^m  {{\dot q_k}{z_{kl}}{\dot q_l}}
\end{equation}
Suppose we have a principal singular vector ${{{\mathbf{q'}}}}$ with negative elements. Let positions of these negative elements be a set $S=\{{k|\mathbf q'_k<0}\}$. We always can construct a new $m$-dimensional vector $\mathbf h$ whose $k$-th element be $q'_k$ if $k\in S$, be $-q'_k$ otherwise. we can find $\mathbf h$ would be non-negative and have:
\begin{equation}
\sum\limits_{k=1}^m \sum\limits_{l=1}^m {{h_k}{z_{kl}}{h_l}} \ge \sum\limits_{k=1}^m \sum\limits_{l=1}^m {{q'_k}{z_{kl}}{q'_l}} =\dot \sigma _1^2
\end{equation}
Thus, we always have a non-negative principal singular vector ${{{\dot{\mathbf q}}_{{1}}}}$. 
Since $\sigma_1{{\dot{\mathbf p}}_1}= {\mathbf{Y}^\top}{{\dot{\mathbf q}}_1} $, the principal singular vector ${\dot{\mathbf p}}_1$ is also non-negative.

Let  $l$ be the ID of the item with the highest popularity. For convenience, here we simply assume $l=1$. Let ${{\mathbf{y}}_k}$ denote the $k$-th column of matrix $\mathbf Y$. Given the non-negative of ${{{\dot{\mathbf p}}_{{1}}}}$ and $y_{ui}\in\{0,1\}$, we have the relation ${{\mathbf{e}}^\top}{\dot{\mathbf p}_1} \ge {{\mathbf{y}}^\top_1}{\dot{\mathbf p}_1}$.
According to ${\mathbf{Y}^\top}{{\dot{\mathbf p}}_1} = \sigma_1{{\dot{\mathbf q}}_1}$, we further have ${{\mathbf{y}}^\top_1}{{\dot{\mathbf p}}_1} = \dot \sigma_1 {\dot q_{11}}$, where $\dot q_{1k}$ denotes the $k$-th element in ${{\dot{\mathbf q}}_1}$. 

Now we turn to derive the lower bound of ${\dot q_{11}}$. Define a matrix $\mathbf B$, which is a mirror of $\mathbf Y$ except the $1$-th column is removed. For any $2\le j \le m$, considering $y_{ui}\in\{0,1\}$, we have:
\begin{equation}
\sum\limits_{i=2}^m {{\mathbf{y}}_j^\top{{\mathbf{y}}_i}}  \le \sum\limits_{i=2}^m {{\mathbf{y}}_i^\top{{\mathbf{y}}_i}}  = \sum\limits_{i=2}^m {{r_i}}  \le {r_{\max }}(\zeta (\alpha ) - 1)
\end{equation}
It means the sum of any row of the matrix $\mathbf B^\top \mathbf B$ is smaller than ${r_{\max }}(\zeta (\alpha ) - 1)$. According to the Perron-Frobenius theorem \cite{pillai2005perron}, we have the largest eigenvalue value of $\mathbf B^\top \mathbf B$ is bounded with: ${\lambda _1}({{\mathbf{B}}^\top}{\mathbf{B}}) \le {r_{\max }}(\zeta (\alpha ) - 1)$. Further, we have:
\begin{equation}
    \sum\limits_{i=2}^m {{{({\dot q_{1i}})}^2}}  = \frac{1}{{\dot \sigma _1^2}}||{{\mathbf{B}}^\top}{{\dot{\mathbf p}}_1}||_2^2 \le \frac{1}{{\dot \sigma _1^2}}{\lambda _{\max }}({{\mathbf{B}}^\top}{\mathbf{B}}) \le \frac{{{r_{\max }}}}{{\dot \sigma _1^2}}(\zeta (\alpha ) - 1)
\end{equation}
Considering $\dot{\mathbf q}_1$ is the normalization term, we can derive the lower bound of the ${{\mathbf{e}}^\top}{{\dot{\mathbf p}}_1}$ as:
\begin{equation}
{{\mathbf{e}}^\top}{{\dot{\mathbf p}}_1} \ge {\dot \sigma _1}{\dot q_{11}} \ge {\dot \sigma _1}\sqrt {1 - \frac{{{r_{\max }}}}{{\dot \sigma _1^2}}(\zeta (\alpha ) - 1)} \label{eq:p3}
\end{equation}
Integrating Eq. (\ref{eq:p3}) into Eq. (\ref{eq:p2}), finally we get the lower bound of the $\cos ({\mathbf{r}},{{\dot{\mathbf q}}_1})$ as:
\begin{equation}
\cos ({\mathbf{r}},{{\dot{\mathbf q}}_1}) \ge \frac{{\dot \sigma _1^2}}{{{r_{\max }}\sqrt {\zeta (2\alpha )} }}\sqrt {1 - \frac{{{r_{\max }}(\zeta (\alpha ) - 1)}}{{\dot \sigma _1^2}}} \label{eq:fi}
\end{equation}
\textbf{Part 4: demonstrating $\dot \sigma _1^2 \ge {r_{\max }}$}. Let define a matrix ${\mathbf{Z }}={{\mathbf{Y}}^\top}{\mathbf{Y}}$. Let $l$ be the ID of the item with the highest popularity. It is easily to find that the $ll$-th element in ${\mathbf{Z }}$ have ${z_{ll}}=r_{max}$. Let ${\mathbf{v}}$ be a one-hot vector whose $l$-th element is one. we have:
\begin{equation}
{r_{\max }} = {{\mathbf{v}}^\top}{\mathbf{Zv}} \le \mathop {\max }\limits_{||{\dot{\mathbf q}}|| = 1} {{\dot{\mathbf q}}^\top}{\mathbf{Z}\dot {\mathbf{q} }} = \dot \sigma _1^2
\end{equation}
Given $\alpha>2$, we have ${\zeta (\alpha )}\le 2$. Considering $\dot \sigma _1^2 \ge {r_{\max }}$, Eq. (\ref{eq:fi}) can be further bounded with: 
\begin{equation}
\cos ({\mathbf{r}},{{\dot{\mathbf q}}_1}) \ge \sqrt {\frac{{2 - \zeta (\alpha )}}{{\zeta (2\alpha )}}}
\end{equation}

Due to the alignment of the principal singular values and vectors of $\mathbf{Y}$ and $\mathbf{ \hat Y}$, we have:
\begin{equation}
\cos ({\mathbf{r}},{{\mathbf{ q}}_1}) \ge \sqrt {\frac{{2 - \zeta (\alpha )}}{{\zeta (2\alpha )}}}
\end{equation}

\subsection{Proof of Theorem 2}
\label{appendix: thm2}
Let $S$ be a set of users where the most popular item occupies the top-1 recommendation. Let $l$ be the ID of the most popular item. $S$ can be wrriten as:
\begin{equation}
S = \{ u \in \mathcal U|{\hat{y}_{ul}} > {\hat{y}_{ui}},\forall i \in \mathcal{I}/l\} 
\end{equation}
The ratio of the most popular item occupying top-1 recommendation can be written as $\eta  = |S|/n$. We then do some transformation of the condition: 

\begin{equation}
\begin{aligned}
   {\hat{y}_{ul}}  &> {\hat{y}_{ui}},\forall i \in \mathcal{I}/l\\
     \Leftrightarrow\sum\limits_{k=1}^L {{\sigma _k}{p_{ku}}{q_{kl}}}  & > \sum\limits_{k=1}^L {{\sigma _k}{p_{ku}}{q_{ki}}} ,\forall i \in \mathcal{I}/l\\
     \Leftrightarrow {\sigma _1}{p_{1u}}({q_{1l}} - {q_{1i}}) & > \sum\limits_{k=2}^L {{\sigma _k}{p_{ku}}({q_{ki}} - {q_{kl}})} ,\forall i \in \mathcal{I}/l \label{eq:con}
\end{aligned}
    \end{equation}
Given the alignment of $\mathbf r$ and $\mathbf q_1$, and the pow-law distribution of the popularity, for any item $i \in \mathcal{I}/l$, the l.h.s of Eq. (\ref{eq:con}) can be bounded by:
\begin{equation}
{\sigma _1}{p_{1u}}({q_{1l}} - {q_{1i}}) = {\sigma _1}{p_{1u}}\frac{{{r_l} - {r_i}}}{{||{\mathbf{r}}||}} \ge {\sigma _1}{p_{1u}}\frac{{1 - {2^{ - \alpha }}}}{{\sqrt {\zeta (2\alpha )} }}
\end{equation}
Besides, given the normalization of the singular vectors, for any item $i \in \mathcal{I}/l$, we can bound the r.h.s of Eq. (\ref{eq:con}) as:
\begin{equation}
\sum\limits_{k=2}^L {{\sigma _k}{p_{ku}}({q_{ki}} - {q_{kl}})}  \le \sum\limits_{k=2}^L{\sqrt 2 {\sigma _k}}
\end{equation}
due to the fact that ${{p_{ku}}}\le 1$ and: 
\begin{equation}
{({q_{ki}} - {q_{kl}})^2} = q_{ki}^2 + q_{kl}^2 - 2{q_{ki}}{q_{kl}} \le 2(q_{ki}^2 + q_{kl}^2) \le 2
\end{equation}
Thus, the condition ${y_{ul}} > {y_{ui}},\forall i \in \mathcal{I}/l$ holds if the following inequality holds:
\begin{equation}
{\sigma _1}{p_{1u}}\frac{{1 - {2^{ - \alpha }}}}{{\sqrt {\zeta (2\alpha )} }} > \sum\limits_{k=2}^L{\sqrt 2 {\sigma _k}}
\end{equation}
It means that we have the lower bound of the $\eta$ as:
\begin{equation}
    \begin{aligned}
    \eta  &= \frac{1}{n}|\{ u \in U|{y_{ul}} > {y_{ui}},\forall i \in \mathcal{I}/l\} |\\
    & \ge \frac{1}{n}|\{ u \in U|{\sigma _1}{p_{1u}}\frac{{1 - {2^{ - \alpha }}}}{{\sqrt {\zeta (2\alpha )} }} > \sum\limits_{k=2}^L {\sqrt 2 {\sigma _k}} \} |\\
     &= \frac{1}{n}|\{ u \in U|{p_{1u}} > \frac{{\sqrt {2\zeta (2\alpha )} }}{{1 - {2^{ - \alpha }}}}(\frac{{\sum\limits_{k=1}^L {{\sigma _k}} }}{{{\sigma _1}}} - 1)\} |\\
     &= \frac{1}{n}\phi (\frac{{\sqrt {2\zeta (2\alpha )} }}{{1 - {2^{ - \alpha }}}}(\frac{{\sum\limits_{k=1}^L {{\sigma _k}} }}{{{\sigma _1}}} - 1))
    \end{aligned}
\end{equation}
where $\phi(a)=\sum_{u\in \mathcal U} {\mathbf I[p_{1u}> a]}$ is an inverse cumulative function calculating the number of elements $p_{1u}$ in the left principal singular vector $\mathbf p_{1}$ exceeding a given value $a$, and the function $\mathbf I[.]$ signifies an indicator function.

\subsection{Theoretical Analysis on Dimension Collapse via Gradient Optimization}
\label{appendix:gradient}
Here we begin by invoking the gradient dynamic theorem from \cite{chou2023gradient,saxe2019mathematical,arora2019implicit} to elucidate the dimension collapse phenomenon and then develop a theorem to illustrate how the singular values impact the popularity bias in recommendations.

\begin{theorem}[Trajectory of singular values (Eq. (6) in \cite{saxe2019mathematical})]
  When training an MF model via gradient flow (gradient descent with infinitesimally small learning rate), \ie  $\frac{d}{{dt}}{\mathbf{U}}(t) =  - \frac{{dL}}{{d{\mathbf{U}}}}$, $\frac{d}{{dt}}{\mathbf{V}}(t) =  - \frac{{dL}}{{d{\mathbf{V}}}}$,  the trajectory of singular values during the learning process obeys:
  \begin{equation}
    \sigma_k(t) = \frac{s_k e^{2s_kt}}{e^{2s_kt}-1+s_k/\sigma_k(0)} \label{eq3}
  \end{equation}
  where $s_k$ signifies the terminal value of the $k$-th singular value, \ie $\sigma_k(t) \to s_k$ as $t \to \infty$.
  \end{theorem}
This theorem illustrates a sigmoidal trajectory that begins at some initial value $\sigma_k(0)$ at time $t=0$ and rises to $s_k$. The growing trajectory of singular values depends on their respective convergence values. It is coincident with the phenomenon presented in Figure \ref{e-c} --- \ie larger singular values are prioritized. Those small singular values require much more time to reach optimum, easily resulting in dimension collapse.

\section{Additional Experiments}

\subsection{Long-tailed Distribution and Bias Amplification in Recommendations}\label{appendix: bias amplify}

\begin{figure}[t]
    \centering
    \includegraphics[width=0.99\linewidth]{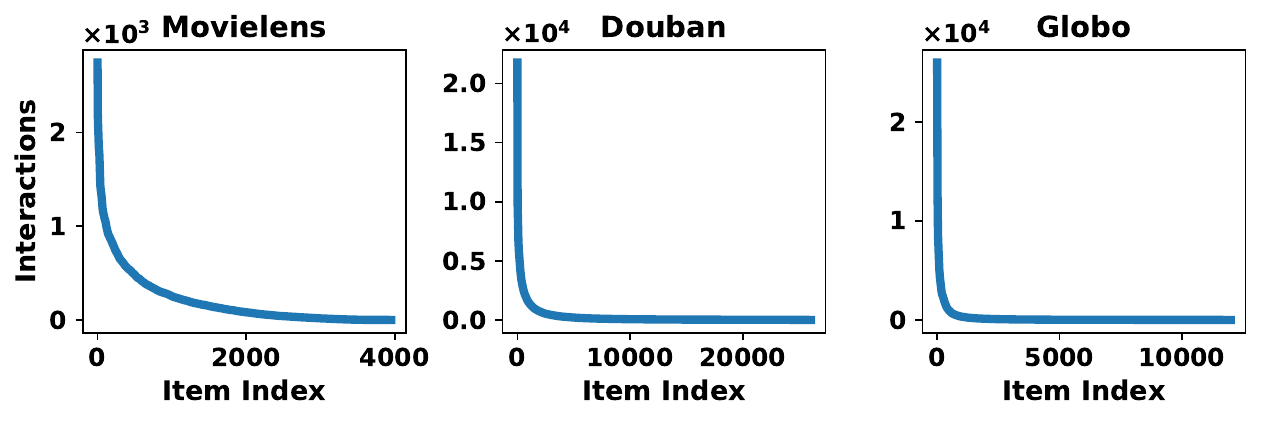}
    \caption{Long-tailed distribution of item popularity in recommendation datasets.}
    \label{Itempop}
\end{figure}

Figure \ref{Itempop} shows the distribution of item popularity (the number of interactions of an item) in the three benchmark datasets. It presents a significant long-tail distribution: a small portion of popular items at the head have a high number of interactions, while the majority of items in the tail have very few interactions. Table \ref{tab:ratio_percentage} shows the proportion of interactions of the top 20\% popular items to all interactions. In all three datasets, the interactions of the top 20\% popular items accounted for over 60\% of all interactions, and in the Globo dataset, it even exceeded 90\%.

\begin{figure}[t]
    \centering
    % \hspace{-5mm}
    \includegraphics[width=0.99\linewidth]{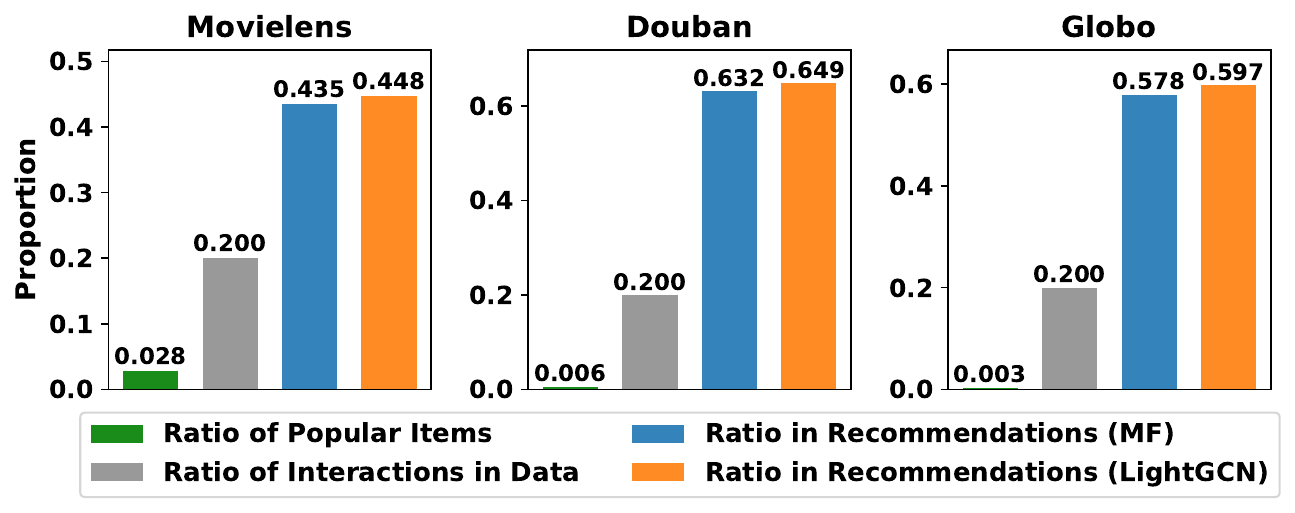}
    \caption{Illustration of popular bias amplification: We divide items into five groups according to their popularity as recent work \cite{zhang2021causal}, and focus on the most popular group. The chart displays four bars, representing the ratio of the items in the most popular group, the percentage of interactions originating from these items in the training set, and the percentages of these items appearing in recommendations from MF and LightGCN, respectively.}
    \label{Rec_proportion}
\end{figure}

Figure \ref{Rec_proportion} shows the popularity bias amplification effect in the three benchmark datasets. In all three datasets, a mere 3\% of the most popular items accounting for 20\% of total interactions occupy over 40\% recomendation slots, and in Douban dataset, it even reaches 60\%. The disparity in the proportion of interactions to recommendation results effectively demonstrate the amplification effect of popularity bias.

% \begin{table}[t]\small
%     \centering
%     \caption{Comparison table between the actual spectral norm and the estimated approximation.}
%     \renewcommand\arraystretch{1.25}
%     \begin{tabular}{c|cc|cc|cc}
%     \toprule 
%      \multirow{2}{*}{\textbf{Model}}&\multicolumn{2}{c}{\textbf{Movielens-1M}}\vline&\multicolumn{2}{c}{\textbf{Douban}}\vline&\multicolumn{2}{c}{\textbf{Globo}}\\
%      \cline{2-7}&$||{\mathbf{\hat Y}}||^2$&$||{\mathbf{\hat Y}}{{\mathbf{\tilde q}}_1}||^2$&$||{\mathbf{\hat Y}}||^2$&$||{\mathbf{\hat Y}}{{\mathbf{\tilde q}}_1}||^2$&$||{\mathbf{\hat Y}}||^2$&$||{\mathbf{\hat Y}}{{\mathbf{\tilde q}}_1}||^2$\\
%      \midrule
%      MF+MSE &5.627\times 10^5 &5.613 \times 10^5 &1.160\times 10^7 &1.155 \times10^7&8.321\times 10^6 &8.309\times 10^6\\
%      MF+BCE &5.629\times 10^5 &5.620 \times 10^5&1.161\times 10^7 &1.157 \times10^7&8.327\times 10^6&8.316\times 10^6\\
%     \bottomrule
%     \end{tabular}
%     \label{sigma}
% \end{table}


\begin{table}[t]
    \centering
    \caption{The proportion of interactions of the top 20\% of popular items in the total number of interactions.}
    \begin{tabular}{cccc}
    \toprule 
     &\textbf{Movielens-1M}&\textbf{Douban}&\textbf{Globo}\\
     \midrule
     \% &67.3&86.3&90.5\\
    \bottomrule
    \label{tab:ratio_percentage}
    \end{tabular}
\end{table}
\begin{table}[t]
    \centering
    \small
    \caption{The value of $\sigma_1^2$ and $r_{max}$ in different datasets and recommendation models.}
    \renewcommand\arraystretch{1.35}
    \tabcolsep=2pt
    \begin{tabular}{ccccccc}
    \toprule 
     \multirow{2}{*}{\textbf{Model}}&\multicolumn{2}{c}{\textbf{Movielens}}&\multicolumn{2}{c}{\textbf{Douban}}&\multicolumn{2}{c}{\textbf{Globo}}\\
     \cmidrule(r){2-3}\cmidrule(r){4-5}\cmidrule(r){6-7}&$\sigma_1^2$&$r_{max}$&$\sigma_1^2$&$r_{max}$&$\sigma_1^2$&$r_{max}$\\
     \midrule
     \textbf{MF} &$5.6\times 10^5$ & $4.6\times10^3$&$1.2\times 10^ 7$&$4.7\times10^4$&$8.3\times 10^6$&$5.0\times 10^4$\\
     \textbf{LightGCN} &$5.7\times 10^5$&$4.6\times10^3$&$1.2\times 10^ 7$&$4.9\times 10^4 $&$8.4\times 10^6$&$5.1\times 10^4$\\
     % \textbf{MF} &5.6\times 10^5&4.6\times10^3&1.2\times 10^ 7&4.7\times10^4&8.3\times 10^6&5.0\times 10^4\\
     % \textbf{LightGCN} &5.7\times 10^5&4.6\times10^3&1.2\times 10^ 7&4.9\times 10^4 &8.4\times 10^6&5.1\times 10^4\\
    \bottomrule
    \end{tabular}
    \label{sigma}
\end{table}

\subsection{Comparison between $\sigma_1^2$ and $r_{max}$} \label{appendix:sigma}

Table \ref{sigma} presents the values of $\sigma_1^2$ and $r_{max}$ on three benchmark datasets and different backbone models. It can be observed that in multiple actual datasets, $\sigma_1^2$ is significantly larger than $r_{max}$, exceeding $100$ times and more. Combining the bounds given by Eq.(\ref{eq:t1}), ($\cos({\mathbf{r}},{{\mathbf{q}}_1}) \ge \frac{{\sigma _1^2}}{{{r_{\max }}\sqrt {\zeta (2\alpha )} }}\sqrt {1 - \frac{{{r_{\max }}(\zeta (\alpha ) - 1)}}{{\sigma _1^2}}}$), even if the data is not markedly skewed, \ie $\alpha$ is not very large and $\zeta(\alpha)$ is not very close to 1, there is still a significant similarity between item popularity vector $\mathbf{r}$ and the principal singular vector $\mathbf{q}_1$ due to the considerable ratio between $\sigma_1^2$ and $r_{max}$. 
This observation also helps to explain the prevalence of popularity bias memorization effect in recommendation models and datasets.

% % \subsection{Validation on Approximation}\label{appendix:validatin}
% % We computed the ideal value of $||{\mathbf{U}\mathbf{V}^\top}||_2^2$, as well as the estimated $\frac{||{\mathbf{U}}{{\mathbf{V}}^\top}{\mathbf{V}}{{\mathbf{U}}^\top}{\mathbf{e}}||^2}{{||{\mathbf{U}\mathbf{V}^\top{\mathbf{e}}}||^2}}$ from ReSN. Here we well-training the MF model with two losses on three datasets. The results are shown in the Table \ref{tab:approx}. According to the table, we found that the actual spectral norms and our approximate estimates are very close across diverse losses and datasets.This indicates that the singular vector $\mathbf{\tilde q}_1$ obtained through $\frac{\mathbf{U}\mathbf{V}^\top\mathbf{e}}{||\mathbf{U}\mathbf{V}^\top\mathbf{e}||}$, serves as an accurate surrogate for the true value of $\mathbf{q}_1$.
% % Therefore, the estimated regularization term is a accurate surrogate for the spectral norm $||{\mathbf{U}\mathbf{V}^\top}||^2_2$ which validates the accuracy and rationality of the proposed method.

\begin{figure}[t]
    \centering
    \includegraphics[width=\linewidth]{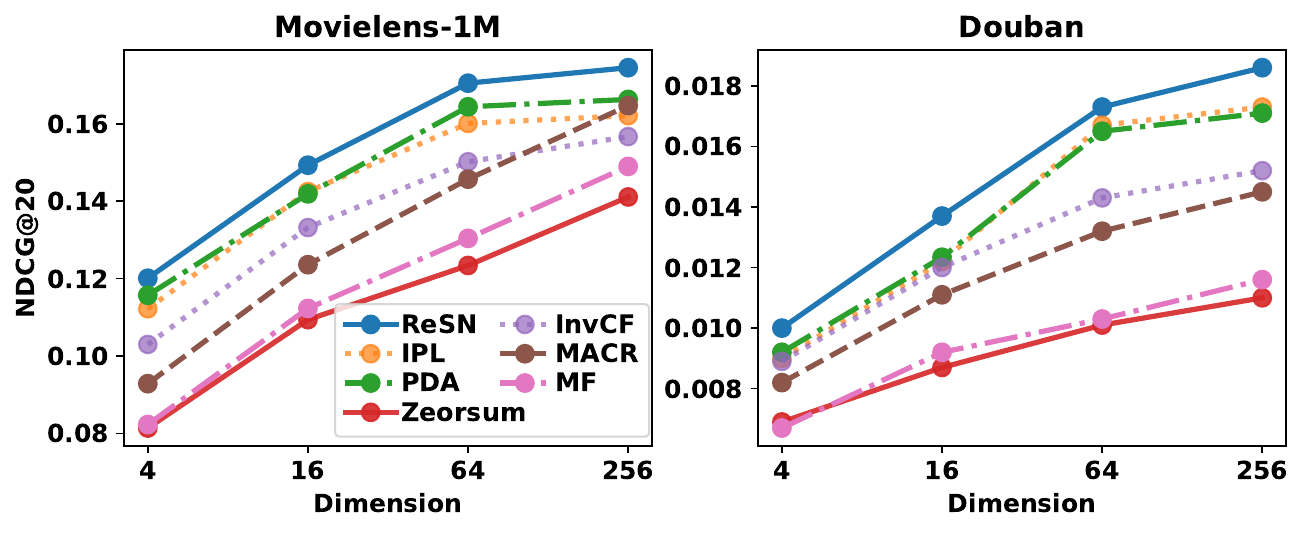}
    \caption{Performance comparison across different embedding dimensions in the Movielens and Douban datasets.}
    \label{dim_performance}
\end{figure}   

\subsection{Performance with Diverse Embedding sizes}\label{appendix:dimension_performance}
To further evaluate the performance of ReSN, we explored the performance of the model under different embedding dimensions, as shown in Figure \ref{dim_performance}. It can be seen that, with the increase of embedding dimensions, the performance of all models gradually improves, and our ReSN can outperform the comparison methods under all different embedding dimensions. This further validates the effectiveness of ReSN in terms of performance.

\begin{table}[t]
    \centering
    \caption{Dataset statistics.}
    \begin{tabular}{lcccc}
    \toprule
        \textbf{Dataset} &\textbf{\#Users}&\textbf{\#Items}&\textbf{\#Interactions}&\textbf{Sparsity} \\ 
        \midrule
        \textbf{Movielens-1M}&6,022&3,043&995,154&5.431\%\\
        \textbf{Douban}&47,890&26,047&7,174,218&0.575\%\\
        \textbf{Globo}&160,377&13,096&2,024,510&0.096\%\\
        \textbf{Gowalla}&29,858&40,981&1,027,370&0.084\%\\
        \textbf{Yelp2018}&31,831&40,841&1,666,869&0.128\%\\
        \textbf{Yahoo!R3}&14,382&1,000&129,748&0.902\%\\
        \textbf{Coat}&290&295&2776&3.24\%\\
        \bottomrule
    \end{tabular}
    
    \label{tab:dataset}
\end{table}
\begin{table*}[t]
    \centering
        \caption{Notations in this paper.}
    \begin{tabular}{ll}
\toprule
        \textbf{Notations} & \textbf{Descriptions} \\
        \midrule
         $u$ & a user in the user set $\mathcal{U}$\\
         $i$ & an item in the item set $\mathcal{I}$\\
         $n$ & the number of users in $\mathcal{U}$\\
         $m$ & the number of items in $\mathcal{I}$\\
         $y_{ui}$& whether user $u$ has interacted with item $i$ \\
         $\mathbf{Y}$& the observed interaction matrix \\
         $r_i$& the number of interactions of item $i$, \ie popularity of item $i$\\
         $\mathbf r$ & the vector of the item popularity over all items \\
         $\mathbf{u}_u$,$\mathbf{v}_i$&embedding vector of user $u$ and item $i$ \\
         $\mathbf{U}$,$\mathbf{V}$ & embedding matrices for all users and items \\
         % $\mathcal{I^\prime }$   & hyper-item space derived from item pairs,\ie $\mathcal{I^ \prime} = \mathcal{I} \times \mathcal{I}$ \\
         % $\mathbf{v}^\prime_{ij}$& the embedding of hyper-items, \ie $\mathbf{v}^\prime _{ij} = \mathbf{v}_i -\mathbf{v}_j$\\
         $\mathbf{\hat{Y}}$ & predicted matrix of all user-item pairs,\ie $\mathbf{\hat{Y}} =\mu(\mathbf{U}\mathbf{V}^\top)$\\
         $\sigma_k$,$\mathbf{p}_k$,$\mathbf{q}_k$& the $k$-th largest singular value and its corresponding left and right singular vector of $\mathbf{\hat{Y}}$\\
         $\alpha$& the shape parameter signifying how severity of long-tail of item popularity\\
         $\zeta(\alpha)$ & Rieman zeta function,\ie $\zeta(\alpha)=\sum_{j=1}^{\infty}\frac{1}{j^{\alpha}}$\\
        $\mathbf{e}$ & a $n$-dimension vector filled with ones\\
    \bottomrule
    \end{tabular}

    \label{tab:notation}
\end{table*}

\section{Experimental Settings}\label{appendix:experiment}
\subsection{Datasets}\label{appendix:dataset}

We adopt seven real-world datasets to evaluate our model:

\begin{itemize}
    \item \textbf{Movielens-1M} \cite{yu2020graph}: Movielens is the widely used dataset from \cite{yu2020graph} and is collected from MovieLens\footnote{https://movielens.org/}. We use the version of 1M. We transform explicit data into implicit feedback by treating all user-item ratings as positive interactions.
    \item \textbf{Douban} \cite{song2019session}: This dataset is collected from a popular review website Douban\footnote{https://www.douban.com/} in China. We transform explicit data into implicit using the same method as applied in Movielens.
    \item \textbf{Globo} \cite{de2018news}: This dataset is a popular dataset collected from the news recommendation website 
 Globo.com\footnote{http://g1.globo.com/}.
    \item \textbf{Yelp2018} \cite{he2020lightgcn} \& \textbf{Gowalla} \cite{he2017neural_a}: Gowalla is the check-in dataset obtained from Gowalla and Yelp2018 is from the 2018 edition of the Yelp challenge, containing Yelp's business reviews and user data. For a fair comparison, these two datasets are used exactly the same as \cite{he2020lightgcn} used.
    \item \textbf{Yahoo!R3} \cite{marlin2009collaborative} \& \textbf{Coat} \cite{schnabel2016recommendations}: These two datasets are obtained from the Yahoo music and Coat shopping recommendation service, respectively. Both datasets contain a  training set of biased rating data collected from normal user interactions and a test set of unbiased rating data containing user ratings on randomly selected items. The rating data are translated to implicit feedback,\ie interactions with ratings larger than 3 are regarded as positive samples.
\end{itemize}

Following the standard 10-core setting, we filter out users and items with less than 10 interactions, and we report the statistics of the above datasets after standardization in Table \ref{tab:dataset}.

\subsection{Implementation Details}\label{appendix:parameter}

We implement ReSN in Tensorflow \cite{abadi2016tensorflow} and the initialization is unified with Xavier \cite{glorot2010understanding}. We optimize all models with Adam \cite{kingma2014adam}. A grid search is conducted to confirm the optimal parameter setting for each model.
To be more specific, learning rate is searched in $\{1e^{-2},1e^{-3},2e^{-4}\}$, weight decay in $\{1e^{-7},1e^{-6},1e^{-5},1e^{-4},1e^{-3}\}$. As for the backbone of LightGCN, we utilize three layers of graph convolution network to obtain the best results,  with or without using dropout to prevent over-fitting.
For ReSN, the coefficient of regularizer $\beta$ is tuned in the range of $\{1e^{-4},1e^{-3},1e^{-2},1e^{-1},5e^{-1},1.0,5.0\}$. 
For compared methods, we closely refer to configurations provided in their respective publications to ensure their optimal performance. 

On the experiments of Pareto curve, besides tuning learning rate and weight decay, we also do following hyperparamter tuning: 1) For PDA, we selected the results of tuning the $\gamma$ and $\tilde {\gamma}$; 2) For MACR, we selected the results of tuning the coefficient $c$; 3) For InvCF, since we found that the differences were not significant after tuning $\alpha$, $\lambda_1$, and $\lambda2$, we reported in the form of a point; 4) For Zerosum, we adjusted its regularization term coefficient, but its results varied greatly and oscillated, so we only reported its best overall performance as a point. 5) For IPL, we selected the results of tuning the regularization term cofficient $\lambda_f$.  All experiments are conducted on a server with Intel(R) Xeon(R) Gold 6254 CPUs.

% \section{Pseudo-Code For ReSN}\label{appendix:pseudo}
% Algorithm \ref{algor:resn} show the learning algorithm of ReSN. The pseudo code illustrates the key procedures of ReSN.
% For a detailed explanation of each step please refer to the main text.

% \begin{algorithm}
% \caption{ReSN's main learning algorithm.}
% \KwIn{Number of users and items: $n$,$m$; user-item interaction data ${\mathbf{Y}}$; weight of the regularization term $\beta$; the dimension size $d$; other hyperparameters}
% \KwOut{Predicted score ${\mathbf{\hat Y}}$}
% Randomly initialize all parameters

% \For{ epoch \in \{1,...,MaxIters\} }{
%     \For{each minibatch (u,i)}{
%         Get user and item embeddings {\mathbf{U}}, {\mathbf{V}}\\
%         Calculate the predicted score $\hat{y}_{ui} = \mu(\mathbf{u}_u\mathbf{v}_i^\top)$\\
%         Calculate the recommendation loss ${\mathcal L_R}(y_{ui},{\hat y_{ui}})$\\
%         Calculate the Spectral Norm regularization term ${\mathcal L_{SN}}=\frac{\beta}{||{\mathbf{V}}{{\mathbf{U}}^\top}{\mathbf{e}}||} ||{\mathbf{U}}{{\mathbf{V}}^\top}{\mathbf{V}}{{\mathbf{U}}^\top}{\mathbf{e}}||^2$\\
%         ${\mathcal {\tilde L}_{{\mathop{\rm Re}\nolimits} SN}} = {\mathcal L_R}((y_{ui},{\hat y_{ui}}) + {\mathcal L_{SN}}$\\
%         Update embedding parameter to minimize ${\mathcal {\tilde L}_{{\mathop{\rm Re}\nolimits} SN}}$
%     }
% }
% \Return{Predicted score ${\mathbf{\hat Y}}$}
% \label{algor:resn}
% \end{algorithm}

\section{Notations}
We summarize the notations used in this paper as follows: uppercase bold letters represent matrices(\eg $\mathbf{Y}$);  lowercase bold letters represent vectors (\eg $\mathbf{r}$); $\Vert \cdot \Vert_2$ to represent the spectral norm of a matrix, \ie the largest singular value of the matrix; and $\Vert \cdot \Vert$ denotes the L2-norm of a vector. Table \ref{tab:notation} provides a more detailed enumeration of the notations used in this paper.
\balance

\end{document}